\documentclass[a4paper,11pt]{article}
\pdfoutput=1 

\usepackage{jcappub} 
\usepackage[normalem]{ulem}
\usepackage{multirow}

\usepackage[utf8]{inputenc}
\usepackage{graphics}
\usepackage{epstopdf}
\usepackage{hyperref}
\usepackage{amsmath}
\usepackage{bigints}
\usepackage{relsize}
\usepackage{psfrag}
\usepackage{epsfig}
\usepackage{subfig}
\usepackage{xcolor}
\usepackage{url}

\usepackage[T1]{fontenc} 
\usepackage[percent]{overpic}
\usepackage{breqn}

\makeatletter
\gdef\@fpheader{}
\makeatother
 \normalsize
\newcommand{\bmat}{\left(\begin{array}}
\newcommand{\emat}{\end{array}\right)}
\newcommand{\be}{\begin{equation}}
\newcommand{\ee}{\end{equation}}
\newcommand{\bea}{\begin{eqnarray}}
\newcommand{\eea}{\end{eqnarray}}

\usepackage{booktabs}
\usepackage{siunitx}

\def\lsim{\raise0.3ex\hbox{$\;<$\kern-0.75em\raise-1.1ex\hbox{$\sim\;$}}}
\def\gsim{\raise0.3ex\hbox{$\;>$\kern-0.75em\raise-1.1ex\hbox{$\sim\;$}}}

\title{\boldmath Induced gravitational waves, metastable cosmic strings and primordial black holes in GUTs}

\author[a]{Rinku Maji,}
\author[b]{Ahmad Moursy,}
\author[c]{and Qaisar Shafi}

\affiliation[a]{Cosmology, Gravity and Astroparticle Physics Group, Center for Theoretical Physics of the Universe,
		  Institute for Basic Science, Daejeon 34126, Republic of Korea}
\affiliation[b]{Department of Basic Sciences, Faculty of Computers and Artificial Intelligence,  \\
Cairo University, Giza 12613, Egypt}
\affiliation[c]{Bartol Research Institute, Department of Physics and Astronomy, \\
	University of Delaware, Newark, DE 19716, USA}
\emailAdd{rinkumaji9792@gmail.com}
\emailAdd{a.moursy@fci-cu.edu.eg}
\abstract{We explore the cosmological and astrophysical implications of a realistic hybrid inflation model based on flipped $SU(5)$. The model contains superheavy metastable cosmic strings arising from a waterfall field that encounters a limited number of $e$-foldings during the inflationary phase. In addition to the gravitational waves emitted by the metastable strings, there also appear scalar induced gravitational waves linked to the waterfall phase transition. These two independent sources of gravitational waves can yield a combined  spectrum that is compatible with the recent PTA measurements, and with additional features that can be probed in future experiments. We also show the appearance of primordial black holes with mass on the order of $10^{29}$ g from the waterfall phase transition, and with an abundance that can be tested in the gravitational lensing experiments.

\vskip 2cm 
{\it \noindent{Dedicated to the memory of our dear friend and collaborator George Lazarides. George was a larger than life figure as well as an outstanding theoretical physicist, and he will be sorely missed.
}}
}
\begin{document}
\maketitle
\flushbottom
\section{Introduction}
\label{sec:intro}
Grand unified theories (GUTs) predict proton decay \cite{Georgi:1974sy,Fritzsch:1974nn,Shafi:1978gg}, and superheavy magnetic monopoles \cite{tHooft:1974kcl,Lazarides:1980cc} that may survive inflation and occur at an observable level \cite{Maji:2022jzu,Moursy:2024hll}. Other topological defects such as domain walls, cosmic strings, intermediate mass monopoles and various composite structures~\cite{Kibble:1982dd,Kibble:1982ae,Vilenkin:1982ks,Lazarides:2019xai,Lazarides:2023iim, Maji:2024pll} can appear in GUTs, depending on the symmetry breaking patterns, with rank greater than four. Detecting such topological defects would be one of the most spectacular confirmation of the physics beyond the Standard Model (BSM).  Topologically stable magnetic monopoles can be present at an observable level provided that their density is diluted during inflation by a suitable number of $e$-foldings in order to satisfy observational constraints from experiments such as MACRO \cite{Ambrosio:2002qq}, IceCube \cite{IceCube:2021eye} and ANTARES \cite{ANTARES:2022zbr}. For example, intermediate scale monopoles in GUTs such as $SO(10)$ should experience $N_M\gtrsim 13$ number of $e$-foldings during inflation~\cite{Maji:2022jzu,Moursy:2024hll}. Superheavy metastable cosmic strings (MSS)~\cite{Antusch:2023zjk,Lazarides:2023rqf,Maji:2023fhv, Ahmed:2023rky,afzal:2023kqs,Antusch:2024nqg,Pallis:2024mip,Ahmed:2024iyd,Maji:2024pll} with a dimensionless string tension parameter $G\mu \sim 10^{-6}$ ($G$ is the Newton's gravitational constant and $\mu$ denotes the string tension) emit a gravitational wave (GW) spectrum that can explain the NANOGrav 15 year \cite{NANOGrav:2023gor, NANOGrav:2023hfp, NANOGrav:2023hvm} and other pulsar timing array data \cite{EPTA:2023fyk,EPTA:2023xxk, Reardon:2023gzh, Xu_2023}. However, to evade the LIGO-VIRGO constraint on the GW spectrum at the decaHertz frequencies \cite{LIGOScientific:2021nrg}, the strings should be partially inflated by about 20-45 $e$-foldings~\cite{Lazarides:2023rqf} or experience an early matter dominated era \cite{Maji:2023fhv,Datta:2024bqp}. Other hybrid topological structures, such as quasistable strings (QSS) and walls bounded by strings (WBS), can emit also gravitational waves compatible with the NANOGrav 15 year results provided that the superheavy strings and monopoles experience an appropriate amount of inflation~\cite{Lazarides:2022jgr, Maji:2023fba,Lazarides:2023ksx,Maji:2024tzg}.

A recent hybrid inflation model based on flipped $SU(5)$~\cite{Lazarides:2023rqf} provides a recipe to entirely inflate the $SU(5)$ superheavy monopoles, but the associated cosmic strings, produced during the waterfall phase, only experience a limited number of $e$-foldings. This kind of intermediate waterfall phase also may produce enhanced primordial curvature perturbations~\cite{Garcia-Bellido:1996mdl,Clesse:2015wea,Braglia:2022phb,Moursy:2024hll}, which can result in the generation of an additional source of GWs, commonly referred to as the scalar induced gravitational waves (SIGW)~\cite{Ananda:2006af,Baumann:2007zm,DeLuca:2020agl,Escriva:2022duf}. Finally, the enhanced scalar perturbations can give rise to primordial black holes (PBHs) which may be present at an observable level~\cite{Escriva:2022duf,Khlopov:2008qy,Belotsky:2014kca}.

In this article, we investigate the possibility that the stochastic gravitational wave background accessible in the pulsar timing array experiments is a superposition of the emission from the two distinct sources above, namely the SIGW and the metastable cosmic string gravitational waves (MSSGW).
We focus on the flipped $SU(5)$ hybrid inflation model~\cite{Lazarides:2023rqf}, where the necessity that the cosmic strings are partially inflated during an intermediate waterfall phase, motivates us to look for another region of the parameter space that allows for the appearance of both SIGWs and MSSGW in explaining the PTA results. The $SU(5)$ symmetry breaking scale is $\approx 5\times 10^{16}$ GeV, and the corresponding proton lifetime is estimated to be of order $10^{37}$ yrs, which is beyond the reach of the Hyper-Kamiokande experiment \cite{Dealtry:2019ldr}. We explore a scenario where the PBHs are produced with an abundance that makes them detectable in the gravitational lensing experiments. Although we focus here on a specific GUT gauge group, we should emphasize that our considerations based on hybrid inflation and waterfall phase transitions can be readily extended to other GUT models. The emission of gravitational waves from two distinct sources and the appearance of primordial black holes is a salient feature of our hybrid inflation model.

The paper is structured as follows. In section \ref{sec:HI-MSS} we briefly review the flipped $SU(5)$ hybrid inflation model in Ref.~\cite{Lazarides:2023rqf} and the production of metastable strings during the waterfall transition. We study the inflationary dynamics and observables in section \ref{sec:inf-dyn}, and present representative benchmark points. In section~\ref{sec:GW}, we discuss the scenario with two distinct sources of GWs and compatibility with the PTA data as well as the LIGO-VIRGO data. In section~\ref{sec:pbh}, we study the production of PBHs during the waterfall phase transition, and our conclusions are summarized in section~\ref{sec:conc}.
%
\section{Metastable cosmic strings from flipped $SU(5)$ model and hybrid inflation}
\label{sec:HI-MSS}
The flipped $SU(5)$ ($SU(5)\times U(1)_X$) model in Ref.~\cite{Lazarides:2023rqf} is an interesting candidate for our study of multiple sources of stochastic gravitational waves background, with the following symmetry breaking chain:\footnote{More details about the model are given in Ref~\cite{Lazarides:2023rqf}.} 
\bea \label{eq:SBpatternX} 
SU(5)\times U(1)_X  &\xrightarrow[]{\left<\Phi\right>} &  SU(3)_c \times SU(2)_L \times U(1)_{Z}  \times U(1)_X \nonumber \\
&\xrightarrow[]{\left<\Psi\right>} &  SU(3)_c \times SU(2)_L \times U(1)_{Y} ,
\eea
where $\Phi \equiv \mathbf{24}_{H}(0)$ and $\Psi \equiv \mathbf{10}_{H}(-1)$. The gauge invariant terms in the scalar potential, realizing the symmetry breaking chain~(\ref{eq:SBpatternX}) and relevant to inflation, are given as follows:
\bea\label{eq:pot_tot}
V &\supset & V_0 - \mu_\Phi^2 \, tr(\Phi^2) -  \dfrac{\mu_1}{3}  \, tr(\Phi^3) + \dfrac{\lambda_1}{4} \, tr(\Phi^4) + \lambda_2 \, [tr(\Phi^2)]^2
- \dfrac{\mu_\Psi^2}{2} \, tr(\Psi^\dagger\Psi) + \dfrac{\lambda_3}{4} \, [tr(\Psi^\dagger\Psi)]^2  \nonumber \\ 
&& + \dfrac{ \lambda_4}{4} \, tr(\Psi^\dagger\Psi\Psi^\dagger\Psi) + \lambda_5 \, tr(\Psi^\dagger \Phi^2 \Psi)+ \lambda_6 \, tr(\Psi^\dagger  \Psi) \, tr(\Phi^2) + \mu_2 \, tr(\Psi^\dagger\Phi\Psi)\nonumber\\
&& +\dfrac{m^2}{2} S^2 
+  \lambda_7 S^2 tr(\Psi^\dagger  \Psi)  -  \lambda_8 S^2 tr(\Phi^2)    \, ,  
\eea
where the adjoint representation $\Phi^\alpha_{\,\, \beta}\equiv \phi_a(T^a)^\alpha_{\, \beta}$, with $T^a$ being the $SU(5)$ generators, and the 10-plet $\Psi$ is a 
$5\times 5$ complex antisymmetric matrix $\Psi^{\alpha\beta}$. The sum over repeated indices is understood, with $a,b,c, \cdots=1,2,\cdots 24$, and $\alpha, \beta, \cdots=1,2,\cdots 5$. The real singlet scalar $S$ represents the inflaton, and we do not include linear, cubic and quartic terms in $S$  assuming that they are adequately small, and hence do not affect the inflation dynamics. The constant vacuum energy $V_0$ is added in order to guarantee a zero cosmological constant in the desired potential minimum.

With the symmetry breaking chain~(\ref{eq:SBpatternX}), $SU(5)$ is broken first via the 24-plet $\Phi$ yielding magnetic monopoles carrying $SU(3)_c$, $SU(2)_L$ and $U(1)_Z$ magnetic fluxes, which are entirely inflated away. The subsequent symmetry breaking $U(1)_ Z \times U(1)_X\to U(1)_Y$ yields cosmic strings which can decay via the quantum-tunneling of monopole-antimonopole pairs~\cite{Lazarides:2023rqf}. The decay width per unit length of the string is expressed as \cite{Preskill:1992ck}
\begin{align}
\Gamma_d=\frac{\mu}{2\pi}\exp{\left(-\pi m_M^2/\mu\right)} ,
\end{align}
with $m_M$ being the monopole mass and $\mu$ the string tension. The superheavy metastable strings can produce gravitational waves compatible with the NANOGrav 15 year data and the constraint from LIGO-VIRGO third run results if: i) the dimensionless string tension parameter $G\mu$ is of order $10^{-6}$ with a metastability factor $\sqrt{\kappa}\equiv m_M/\sqrt{\mu}\sim 8$, and ii) the strings experience a limited number of $e$-foldings before the end of inflation~\cite{Lazarides:2023rqf}.

As explained in Ref.~\cite{Lazarides:2023rqf}, the scalar potential~(\ref{eq:pot_tot}) provides a hybrid inflation model where the Higgs field $\Psi$ plays the role of the waterfall field that is frozen at the origin during inflation until $S$ reaches a critical value $S_c$, at which point the waterfall phase transition is triggered. The Higgs field $\Phi$ is shifted from the origin, following a field dependent minimum during inflation until the time at which $S=S_c$. The inflationary potential during this phase has the form
\bea\label{eq:pot_inf}
V_{\rm inf} &=& V_0 - \dfrac{m_\phi^2}{2} \, \phi^2  + \dfrac{\beta_\phi}{4} \, \phi^4
- \dfrac{m_\psi^2}{2} \, \psi^2 + \dfrac{\beta_\psi}{4} \, \psi^4  + \frac{\beta_{\psi\phi}}{2} \, \psi^2\phi^2 +\dfrac{m^2}{2} S^2 
+  \frac{\beta_{S\psi}}{2} \, S^2 \, \psi^2  -  \frac{\beta_{S\phi}}{2} \,  S^2 \, \phi^2   \, , \nonumber \\
\eea
 where $\phi$ and $\psi$ represent the real canonically normalized components of the scalar fields $\Phi$ and $\Psi$ which acquire the VEVs.  The superrenormalizable terms with coefficients $\mu_1$ and $\mu_2$ are suitably controlled such that their effects on the dynamics are negligible, and for simplicity we set $\mu_1=\mu_2=0$ in Eq.~(\ref{eq:pot_tot}). Also, we assume that all of the remaining coefficients are real. The charged components of $\Phi$ and $\Psi$ are fixed at zero during and after inflation, and the parameters $m_\phi, m_\psi, \beta_\phi, \beta_\psi,\beta_{\psi\phi} , \beta_{S\phi}, \beta_{S\psi}$ can be expressed in terms of the original potential parameters in Eq.~(\ref{eq:pot_tot}). The constant vacuum energy $V_0$ satisfies 
  \bea
 V_0= \dfrac{1}{4} \left(m_\psi^2 v_\psi^2 + m^2_\phi v_\phi^2 \right),
 \eea
where $v_\psi$ and $ v_\phi$ are the vacuum expectation values of $\psi$ and $\phi$ respectively. The inflation trajectory in the $(\psi,\phi)$ plane is given by~\cite{Ibrahim:2022cqs,Lazarides:2023rqf,Moursy:2024hll}
\bea\label{eq:trajec} 
(\psi,\phi) = \left(0 \, ,\,  \sqrt{\frac{m_{\phi }^2+  \beta _{S \phi } S^2 }{\beta _{\phi }}} \right) \,.
 \eea

Th standard hybrid inflation tree level potential \cite{Linde:1993cn} is then modified to a hill-top shaped one~\cite{Ibrahim:2022cqs,Lazarides:2023rqf,Moursy:2024hll}, with the tree level effective potential given by
\bea\label{eq:infpot1}
 V_{\text{inf}}(\widetilde{S})= \widetilde{V}_0\left( 1+  \widetilde{S}^2- \gamma \, \widetilde{S}^4\right),
\eea
where
\bea 
 \widetilde{V}_0 \equiv V_0-\dfrac{m_\phi^4}{4\beta_\phi} 
 \,, \hspace{0.5cm} \widetilde{S} \equiv \sqrt{\dfrac{\eta_0}{2}} \, S
 \,, \hspace{0.5cm} \eta_0 \equiv \frac{m^2 \beta _{\phi }-m_{\phi }^2 \beta _{S\phi }}{\widetilde{V}_0 \,\beta _{\phi }} 
 \,, \hspace{0.5cm} \gamma \equiv \frac{\beta _{S \phi }^2}{\eta _0^2 \widetilde{V}_0 \,\beta _{\phi }} \,,
\eea 
and the critical value of the inflaton field $S_c$, at which the waterfall is triggered is given by
\be 
S_c= \sqrt{\frac{\beta _{\phi } m_{\psi }^2-\beta _{\psi \phi } m_{\phi }^2}{\beta _{\psi \phi } \beta _{{S\phi }}+\beta _{\phi } \beta _{{S\psi }}}} \,.
\ee

The 1-loop Coleman-Weinberg (CW) correction to the tree level inflation potential~(\ref{eq:infpot1}) may yield a significant change to the inflation observables, since the waterfall fields have inflaton dependent masses during inflation. However, it was shown in Ref.\cite{Lazarides:2023rqf} that the CW radiative correction is under control and yields a tiny contribution to the total potential if we introduce an extra pair of fermionic $10_{F_1}$, $\overline{10}_{F_2}$, with $X = 4, -4$ respectively. In our numerical calculations we choose values of the Yukawa couplings $Y_S\, S \, 10_{F_1} \overline{10}_{F_2} $ and $Y_\Phi\, \Phi \, 10_{F_1} \overline{10}_{F_2} $, such that the CW correction is very small compared to the tree level inflation potential~(\ref{eq:infpot1}).

After reaching the global minimum of the scalar potential~(\ref{eq:pot_inf}), the inflaton field $S$ and the waterfall fields $\phi$ and $\psi$ start to oscillate around their respective minima and decay, such that the reheating phase starts. The inflaton field $S$ and the adjoint scalar field $\phi$ decay to the SM Higgs doublets $\cal H$ via the couplings $\delta_1 \, S \, {\cal H}^\dagger {\cal H}$ and $\delta_2 \, \phi \, {\cal H}^\dagger {\cal H} $. On the other hand, the scalar field $\psi$ decay produces right handed neutrinos via the non-renormalizable coupling $\frac{f}{ M_{\rm Pl}} 10_F 10_F 10_H^\dagger \, 10_H^\dagger$. The reheating temperature $T_r$ consistent with the observables is less than or of order $ 10^{12}$ GeV~\cite{Lazarides:2023rqf}.

Our aim is to shed light on representative points in an interesting region of the parameter space with intermediate waterfall phase~\cite{Clesse:2010iz,Kodama:2011vs,Clesse:2015wea,Lazarides:2023rqf,Moursy:2024hll} that continues for $\sim 20-40$ $e$-foldings, with a significant enhancement of the curvature power spectrum from the waterfall phase transition. In this case, SIGWs are generated by second order effects in perturbation theory. Moreover, primordial black holes may be produced with a significant abundance. Moreover, the metastable cosmic strings produced during the waterfall phase are partially inflated, and they constitute another source of gravitational waves (MSSGW) in addition to the SIGWs. We explore the interplay between the two sources of gravitational waves and the potential production of PBHs in the next sections.
%
\section{Inflationary dynamics and observables}
\label{sec:inf-dyn}
\begin{figure}[htbp!]
    \centering
    \includegraphics[width=0.45\linewidth]{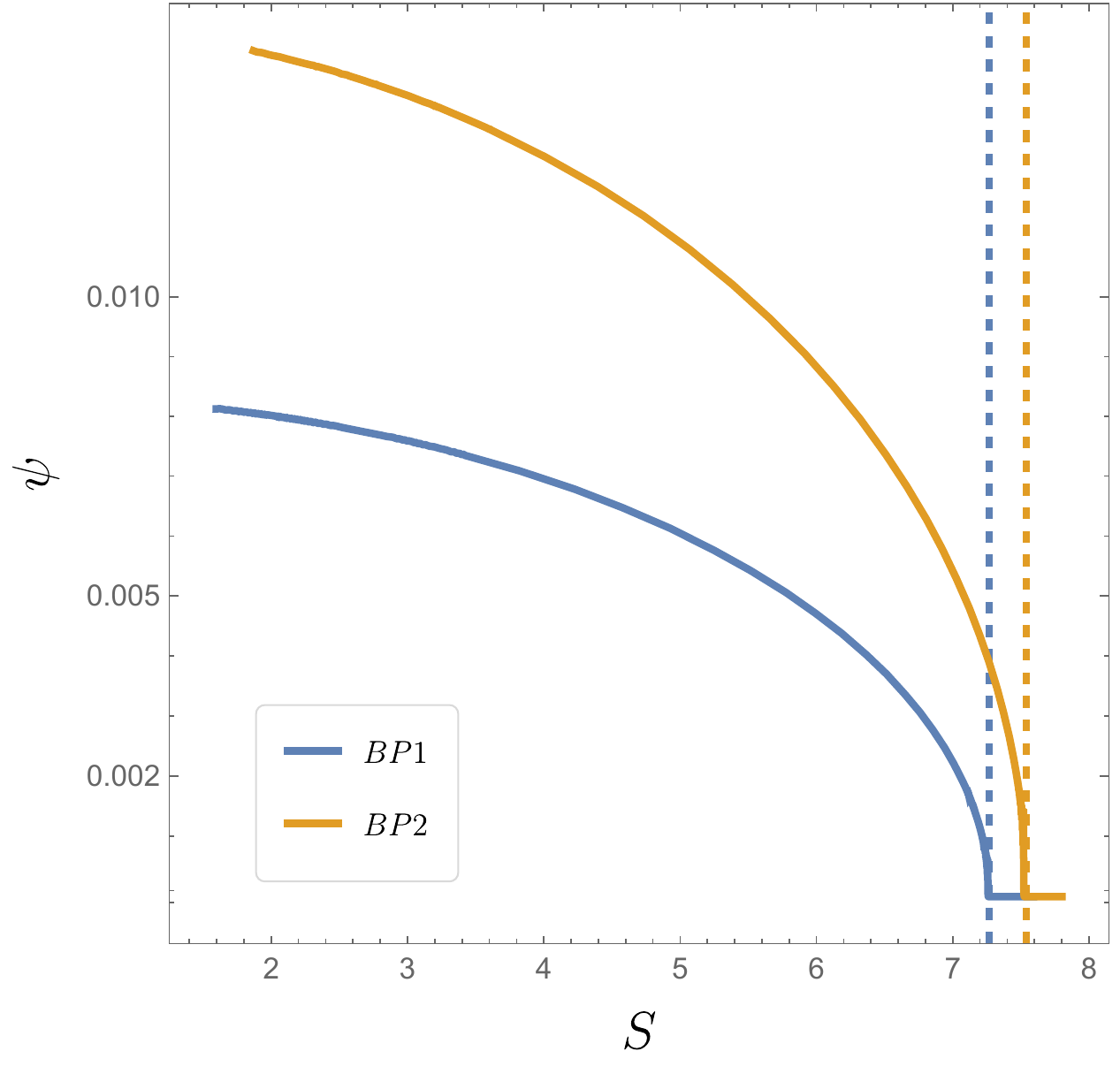}
    \hspace{0.5cm}
    \includegraphics[width=0.45\linewidth]{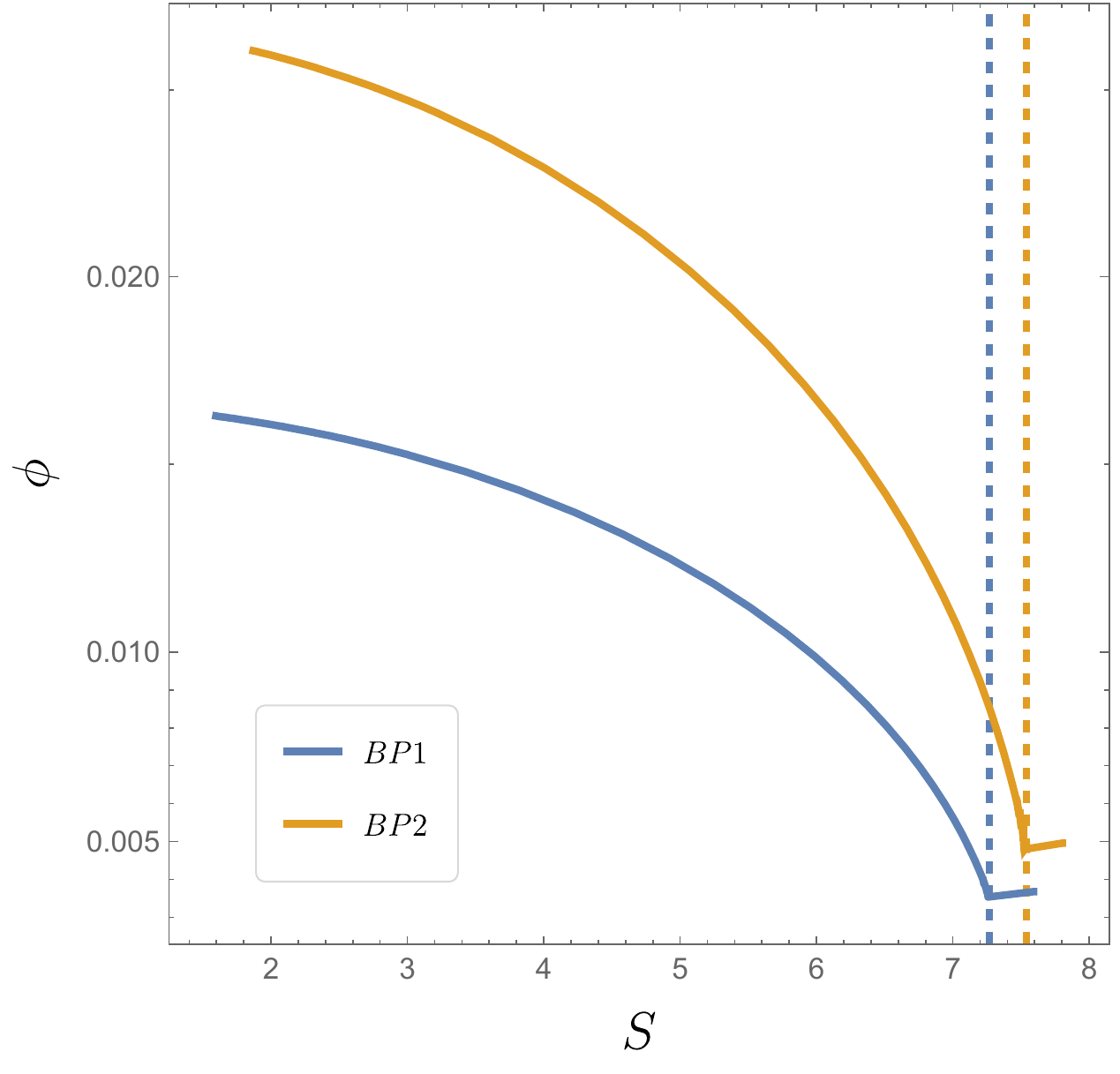}
    \caption{\label{fig:traj} The inflationary trajectory, represented by the solid curves, in the $S-\psi$ plane (left panel), and $S-\phi$ plane (right panel) for the two benchmark points in Table~\ref{tab:merged}. The vertical dashed lines correspond to the critical value $S_c$ for BP1 and BP2.}
\end{figure}
We define the Hubble slow-roll parameters as follows
\bea 
\epsilon_H &\equiv&  \dfrac{1}{2H^2} \sum_{n=1}^{3} {\dot{\varphi}_n}^2 =
\dfrac{1}{2} \sum_{n=1}^{3} {\varphi_n'}^2\,, \,\,\,\,\,
 \eta_H \equiv  \dfrac{\epsilon_H'}{ \epsilon_H}  \,,
 \eea
with the Hubble parameter $H$ given by
\bea\label{eq:FLeq}
H^2=\left( \frac{\dot{a}}{a} \right)^2=\dfrac{1}{3M^2_{\rm Pl}} \left[\dfrac{1}{2} \sum_{n=1}^{3} \dot{\varphi}_n^2 +V(\varphi_n) \right],
\eea \label{eq:KGt}
where $a$ is the scale factor. Here, a dot denotes the derivative with respect to the cosmic time $t$, and a prime denotes the derivative with respect to the number of $e$-foldings variable $N$. The classical field dynamics is governed by the Klein-Gordon equations which can be recast in the form 
\bea\label{eq:Neqm}
\varphi_n''+ \left( 3 -\epsilon_H \right)  \varphi_n' + \left( 3 -\epsilon_H \right)  \dfrac{V_n}{V} = 0\,,
\eea
where $n$ stands for $S, \psi ,\phi$, and $V_n$ is the derivative of the potential with respect to $\varphi_n$. Figure~\ref{fig:traj} depicts the inflationary trajectory in the $S-\psi$ and $S-\phi$ planes, for the benchmark points in Table~\ref{tab:merged}.

Denoting the Bardeen potentials by $\Phi_{\text{B}}$ and $\Psi_{\text{B}}$, the perturbed metric takes the form 
\begin{equation}
ds^2=a (\tau) ^2 \left[ -(1+2 \Phi_{\text{B}})d \tau^2+  \left(1-2 \Psi_{\text{B}}\right)\delta_{ij} \,  dx^i dx^j \right], 
\label{eq.metr}
\end{equation}
where $\tau$ is the conformal time defined by $d t\equiv a d \tau $. The evolution of the scalar field perturbations $\delta \varphi_n$ and $\Phi_{\text{B}}$ is described by the following equations~\cite{Ringeval:2007am,Clesse:2013jra,Moursy:2024hll} 
\bea
	\delta \varphi_n^{''}+(3-\epsilon_H)\delta \varphi_n^{'}+\dfrac{1}{H^2} \sum_{m=1}^{3}V_{n m}\delta \varphi_m+\dfrac{k^2}{a^2H^2}\delta \varphi_n &= & 4\Phi_{\text{B}}^{'}\,\varphi'_n-\dfrac{2\,\Phi_{\text{B}}}{H^2}V_n, \\
	\Phi^{''}_{\text{B}}+(7-\epsilon_H)\,\Phi^{'}_{\text{B}}+\left(2\dfrac{V}{H^2}+\dfrac{k^2}{a^2H^2}\right)\Phi_{\text{B}}  & = &  - \dfrac{1}{H^2} \sum_{m=1}^{3} V_m\,\delta \varphi_m,
\eea 
where $k$ is the co-moving wave vector, and the initial conditions at $N=N_{\rm ic}$ are given by~\cite{Ringeval:2007am,Clesse:2013jra,Moursy:2024hll}:
\bea \label{eq:IC}
	\delta \varphi_{n}(k,N_{\rm ic}) &=& \dfrac{1}{a \,  \sqrt{2 k} }, \\
	\delta \varphi_{n}^{'} (k,N_{\rm ic})  &=& - \dfrac{1}{ a \, \sqrt{2 k} } \left(1+ i \dfrac{k}{a \, H}\right), \\
\Phi_{\text{B}}(k,N_{\rm ic}) &=& \dfrac{1}{2 \left(\epsilon_H - \dfrac{k^2}{a^2 \, H^2 }\right)}
\mathlarger{\mathlarger{\sum}}_{m=1}^{3} \left(  \varphi_m^{'}\delta \varphi_m^{'}+   3 \varphi_m^{'}\delta \varphi_m + \dfrac{1}{H^2} V_m  \, \delta\varphi_m \right) , \\
	\Phi_{\text{B}}^{'}(k,N_{\rm ic}) &=& \mathlarger{\mathlarger{\sum}}_{m=1}^{3}\dfrac{1}{2} \varphi_m^{'}\delta \varphi_m-\Phi_{\text{B}} \, .
\eea
The scalar power spectrum $P_\zeta(k)$ is then given by \cite{Ringeval:2007am,Clesse:2013jra,Moursy:2024hll}
\bea
\label{eq:PR}
	P_\zeta(k)=\dfrac{k^3}{2\pi^2}\left|\Phi_{\rm B}+\dfrac{\sum\limits_{m=1}^{3} \varphi'_m\delta \varphi_m}{\sum\limits_{m=1}^{3}\varphi^{'2}_m}\right|^2.
\eea
\begin{table}[h!]
 \centering
 \begin{tabular}{| >{\centering\arraybackslash}p{3.0cm} |c | c | c | c |}
 \hline 
 \multicolumn{2}{|c|}{Parameters and observables}  & BP1 & BP2  \\
 \hline 
 \multirow{10}{=}{{Potential parameters}} & $\widetilde{V}_0[M_{\rm Pl}^4]$ & $1.982 \times 10^{-11}$ & $1.96 \times 10^{-11}$  \\
& $\eta_0[M_{\rm Pl}^{-2}]$ & $0.013$ & $0.013$  \\
& $\gamma$ & $1.2$ & $1.15$  \\
& $m$ [GeV] & $1.30\times 10^{12}$ & $1.27 \times 10^{12}$ \\
& $m_\psi$ [GeV] & $1.71\times 10^{15}$ & $9.55\times 10^{14}$ \\
& $m_\phi$ [GeV] & $9.64\times 10^{14}$ & $6.025\times 10^{14}$ \\
& $\beta_\psi$ & $1.52$ & $0.326$ \\
& $\beta_\phi$ & $0.095$ & $0.028$ \\
& $\beta_{\psi\phi}$ & $-0.379$ & $-0.096$ \\
& $\beta_{S\psi}$ & $9.94\times 10^{-8}$ & $4.164 \times 10^{-8}$ \\
& $\beta_{S\phi}$ & $1.96\times 10^{-8}$ & $1.038 \times 10^{-8}$ \\
\hline\hline
\multirow{5}{=}{VEVs and physical masses of scalar fields in GeV} & $v_\phi$  & $4\times 10^{16}$ & $6.45 \times  10^{16}$ \\
& $v_\psi$  & $2\times 10^{16}$ & $3.5\times 10^{16}$ \\
& $M_{\psi'}$  & $1.62\times 10^{15}$ & $9.836 \times 10^{14}$ \\
& $M_{\phi'}$  & $3.89 \times 10^{16}$ & $3.215 \times 10^{16}$ \\
& $M_S$  & $3.18 \times 10^{12}$ & $3.067\times 10^{12}$ \\
\hline\hline
\multirow{5}{=}{Inflation observables, $e$-foldings and inflaton value} & $A_s$ & $2.070\times 10^{-9}$ & $2.041\times 10^{-9}$ \\
& $n_s$ & $0.963$ & $0.964$ \\
& $r$ & $7.4\times 10^{-4}$ & $6.42\times 10^{-4}$ \\
& $\Delta N_*$ & $53.07$ & $53.24$ \\
& $\Delta N_c$ & $32.77$ & $35.90$ \\
& $S_*[M_{\rm Pl}]$ & $7.505$ & $7.802$ \\
\hline\hline
\multirow{3}{=}{Cosmic times in seconds} & $t_c$ & $1.96 \times 10^{-36}$ & $1.675 \times 10^{-36}$ \\
& $t_e$ & $5.73\times 10^{-36}$ & $5.81\times 10^{-36}$ \\
& $t_r$ & $4.18\times 10^{-29}$ & $1.78\times 10^{-29}$ \\
\hline\hline
\multirow{2}{=}{Equation of state parameter} & \multirow{2}{*}{$\omega_r$} & \multirow{2}{*}{$0$} & \multirow{2}{*}{$0$} \\
&&& \\
\hline 
 \end{tabular}
 \caption{ Two benchmark points (BP1 and BP2) satisfying the CMB data and accounting for both SIGW and MSSGW.  The benchmark points include the parameters of the potentials in Eqs.~\eqref{eq:pot_inf} and \eqref{eq:infpot1}, VEVs and physical masses of the inflaton and waterfall fields, CMB observables, the number of $e$-foldings during inflation, $\Delta N_*$ starting from the pivot scale ($S=S_*$) and $\Delta N_c$ starting from the beginning of waterfall ($S=S_c$). $t_c$, $t_e$, and $t_r$ denote the cosmic times when $S=S_c$, at the end of inflation, and at reheating, respectively.}
 \label{tab:merged}
\end{table}
In our numerical simulations, integrating the perturbation equations and calculation of the primordial power spectrum, we use the method described in~\cite{Ringeval:2007am,Clesse:2013jra,Clesse:2015wea,Moursy:2024hll}. In addition, we assume an initial displacement $\psi_0\sim H/(2\pi)$ at the time when $S=S_c$~\cite{Kodama:2011vs,Clesse:2015wea,Moursy:2024hll}. 
The power spectrum is normalized at the pivot scale $k_*=0.05 \, {\rm Mpc}^{-1}$ to satisfy the Planck constraints $P_\zeta(k_*)= (2.099\pm0.101) \times 10^{-9}$~\cite{BICEP:2021xfz,Planck:2018jri}. The total number of $e$-foldings during inflation is calculated, taking into account the thermal history of the universe, from the relation \cite{Liddle:2003as,Chakrabortty:2020otp,Kawai:2023dac}: 
\begin{equation}\label{eq:Ns2}
\Delta N_* \simeq 61.5 + \frac{1}{2} \mathrm{ln} \frac{\rho_*}{M_{\rm Pl}^4}-\frac{1}{3(1+\omega_r)} \mathrm{ln} \frac{\rho_e}{M_{\rm Pl}^4} + \left(\frac{1}{3(1+\omega_r)} - \frac{1}{4} \right)\mathrm{ln} \frac{\rho_r}{M_{\rm Pl}^4} \ ,
\end{equation}
where $M_{\rm Pl}\approx 2.44\times 10^{18}$ GeV is the reduced Planck mass, $\rho_* = V(\widetilde{S}_*)$  stands for the energy density of the universe when the pivot scale exits the horizon, $\rho_e = V(\widetilde{S}_e)$ denotes the energy density at the end of inflation. The energy density at the reheating time is given by $\rho_r = (\pi^2/30) g_*T_r^4$, with $g_*= 106.75$ being the effective number of massless degrees of freedom, corresponding to the SM spectrum. The quantity $w_r$ denotes the effective equation-of-state parameter from the end of inflation until reheating that we set equal to zero, since the inflaton potential is quadratic around its true minimum~\cite{Senoguz:2015lba,Maity:2024odg}. The time at reheating $t_r$ is computed from the relation \cite{Lazarides:1997xr,Lazarides:2001zd}
\be
T_r^2 \approx \sqrt{\dfrac{45}{2\pi^2 \, g_*}} \, \dfrac{M_{\rm Pl}}{t_r}.
\ee 
\begin{figure}[htbp!]
    \centering
    \includegraphics[width=0.8\linewidth]{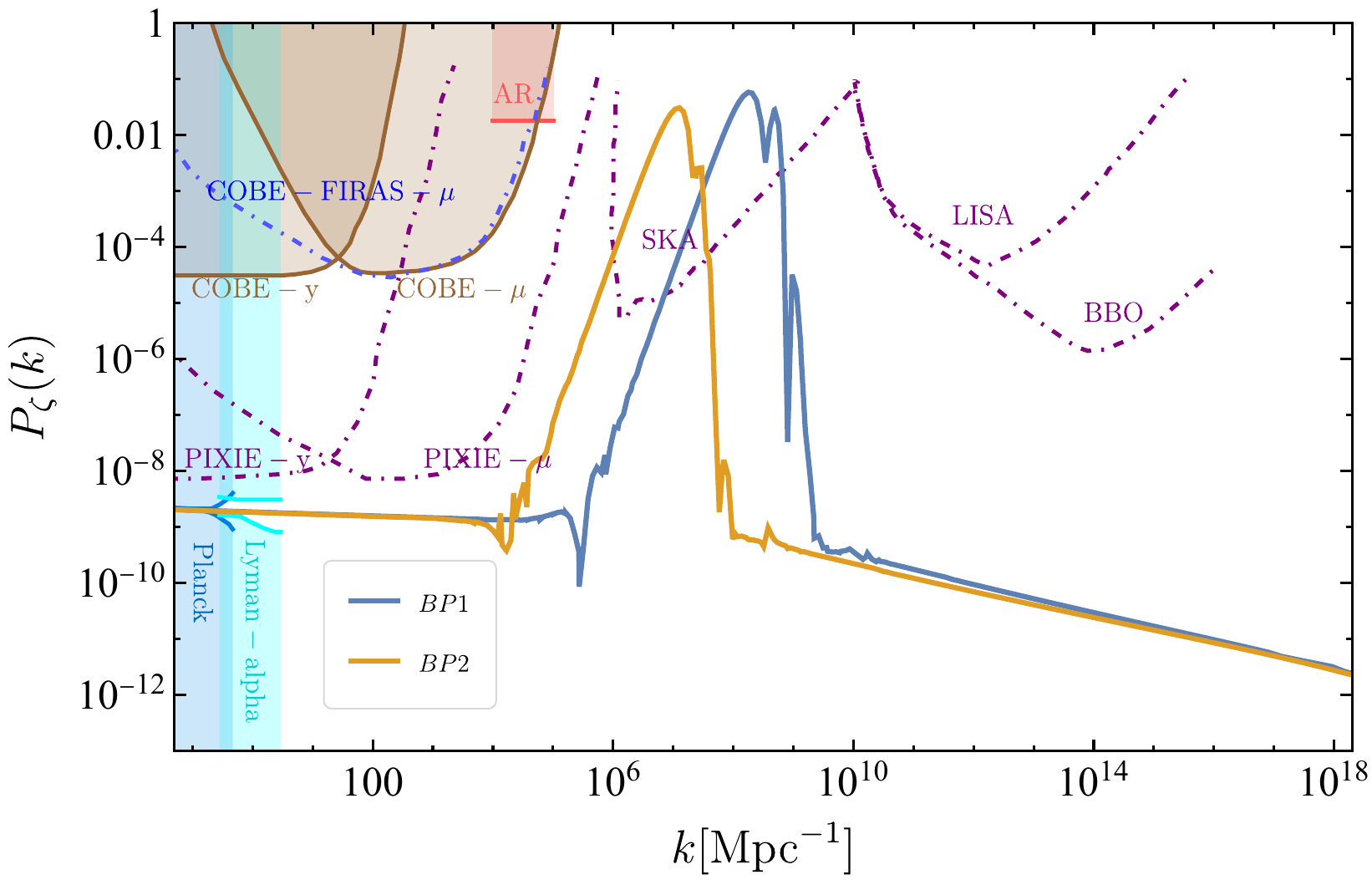}
    \caption{\label{fig:PS} The curvature power spectrum for our model corresponding to benchmark points BP1 and BP2. The shaded regions represent various observational constraints from Planck~\cite{Planck:2018jri}, acoustic reheating (AR)~\cite{Nakama:2014vla} and $\mu$-distortions and $y$-distrortions~\cite{Fixsen:1996nj}. The dot-dashed curves correspond to some future gravitational wave experiments such as LISA, BBO and SKA~\cite{Green:2020jor} as well as the future PIXIE-like detector exploration of $\mu$-distortions and $y$-distrortions~\cite{Kogut:2011xw}.}
\end{figure}
The benchmark points BP1 and BP2 dictated in Table~\ref{tab:merged} correspond to metastable cosmic strings with dimensionless string tension $G\mu \approx 10^{-5}$, and Hubble parameter during the waterfall $H \approx 2.82 \times 10^{-6} \, M_{\rm Pl}$. The curvature power spectrum is enhanced significantly, for both benchmark points, at small scales due to the waterfall phase transition as shown in Figure~\ref{fig:PS}.
\section{Gravitational waves from strings and scalar perturbations}
\label{sec:GW}
As advocated above, the waterfall phase transition plays a dual role in our scenario. First, it is important to partially dilute the cosmic strings density such that the GW spectrum emitted from the sting network is compatible with the LVK bound. Secondly, as stated earlier, it enhances the curvature perturbations at small scales, and hence sources second order tensor perturbations, leading to the scalar induced gravitational waves. Assuming that the scalar induced GWs are produced during the radiation-dominated epoch, the spectrum is calculated from the formula \cite{Lewicki:2021xku,Kohri:2018awv}\footnote{Many detailed studies of SIGW are provided in the literature~\cite{Chatterjee:2017hru,Espinosa:2018eve,Inomata:2019yww,Domenech:2019quo,Domenech:2020kqm,Domenech:2021ztg,Basilakos:2023jvp,Domenech:2024rks,Wang:2019zhj}.}
 \be \label{eq:OmegaGW}
\Omega_{\rm GW}^{\rm SI}h^2 \approx 4.6\times 10^{-4} \left(\frac{g_{*,s}^{4}g_{*}^{-3}}{100}\right)^{\!-\frac13} \!\int_{-1}^1 {\rm d} x \int_1^\infty {\rm d} y \, \mathcal{P}_\zeta\left(\frac{y-x}{2}k\right) \mathcal{P}_\zeta\left(\frac{x+y}{2}k\right) F(x,y) \bigg|_{k = 2\pi f} \,,
\ee
with $g_{*,s}\approx g_{*}$, and the function $F$ is defined as  
\bea
F(x,y) &=&\frac{(x^2\!+\!y^2\!-\!6)^2(x^2-1)^2(y^2-1)^2}{(x-y)^8(x+y)^8} \times \nonumber \\
&&
\!\left\{\left[x^2-y^2+\frac{x^2\!+\!y^2\!-\!6}{2}\ln\left|\frac{y^2-3}{x^2-3}\right|\right]^{\!2} \!+\! \frac{\pi^2(x^2\!+\!y^2\!-\!6)^2}{4}\theta(y-\sqrt{3}) \right\}.
\eea 
%


The gravitational waves from the string network have been extensively studied in the literature \cite{Vachaspati:1984gt,Kibble:1984hp,Vilenkin:2000jqa,Damour:2001bk,Vanchurin:2005pa,Ringeval:2005kr,Olum:2006ix,Leblond:2009fq, Olmez:2010bi,Blanco-Pillado:2013qja,Blanco-Pillado:2017oxo,Cui:2018rwi,Buchmuller:2019gfy,Buchmuller:2021mbb,Dunsky:2021tih,Roshan:2024qnv}. The metastable cosmic string network produces string loops with the number distribution given by \cite{Buchmuller:2021mbb}
\begin{align}\label{eq:source-loop}
n(l,t) =\frac{0.18}{t^{3/2}(l+\Gamma G\mu t)^{5/2}}\begin{cases} \Theta(0.18t-l) & \mathrm{for} \ \  t<t_s \\  e^{-\Gamma_d\left[l(t-t_s)+\frac{1}{2}\Gamma G\mu(t-t_s)^2\right]}\Theta(0.18t_s-\bar{l}) & \mathrm{for} \ \ t>t_s \,\,\,,
\end{cases}
\end{align}
where $\bar{l}=l+\Gamma G\mu (t-t_s)$, $\Gamma\simeq 50$ is a numerical factor, and $t_s=1/\sqrt{\Gamma_d}$.
\begin{figure}[htbp!]
    \centering
    \includegraphics[width=0.8\linewidth]{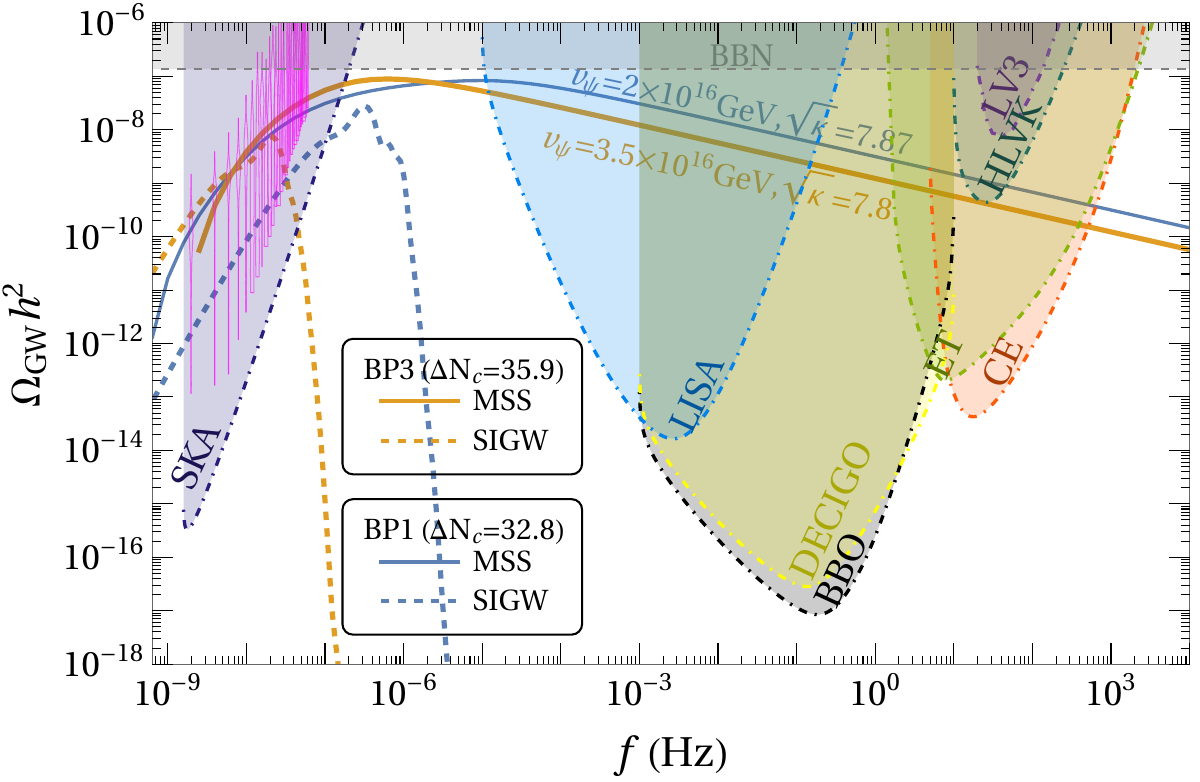}
    \caption{Gravitational waves from the metastable cosmic strings and the associated scalar induced gravitational waves. The gray region depicts the Big Bang Nucleosynthesis (BBN) bound \cite{Mangano:2011ar} on the gravitational wave background for the BP1. The power-law integrated sensitivity curves \cite{Thrane:2013oya, Schmitz:2020syl} for the ongoing and several proposed experiments, namely, HLVK \cite{KAGRA:2013rdx}, CE \cite{Regimbau:2016ike}, ET \cite{Mentasti:2020yyd}, DECIGO \cite{Sato_2017}, BBO \cite{Crowder:2005nr, Corbin:2005ny}, LISA \cite{Bartolo:2016ami, amaroseoane2017laser} and SKA \cite{5136190, Janssen:2014dka} are shown.}
    \label{fig:gws}
\end{figure}
The loops oscillate and radiate gravitational waves. The gravitational wave burst rate per unit spacetime is given by 
\begin{align}
\frac{d^2R}{dz \, dl} = N_B H_0^{-3}\phi_V(z) \frac{2n(l,t)}{l(1+z)}\Delta_B(f,l,z) , 
\end{align}
where $N_B$ is the average number of burst events per oscillation, $H_0$ denotes the Hubble parameter today, $\Delta_B(f,l,z)$ is the observable fraction of the bursts \cite{Damour:2001bk,Leblond:2009fq, Olmez:2010bi}, and 
\begin{align}
\phi_V(z) = \frac{4\pi H_0^3 r^2}{(1+z)^3{H}(z)},
\end{align}
with $r$ denoting the proper distance at redshift $z$. We assume that the cusp events provide the dominant contribution to the gravitational wave background. The wave form of the bursts from a cusp is expressed as \cite{Damour:2001bk},
\begin{align}
h_c(f,l,z) = g_{c}\frac{G\mu  l^{2/3}}{(1+z)^{1/3}r(z)}f^{-4/3},
\end{align}
where $g_{c}\simeq 0.85$ \cite{LIGOScientific:2021nrg}. The gravitational wave background can be expressed as \cite{Leblond:2009fq, Olmez:2010bi}
 \begin{align}\label{eq:GWs}
 \Omega_{\rm GW}^{\rm MSS}(f) = \frac{4\pi^2}{3H_0^2}f^3\int_{z_{*}}^{z_F} dz \int_0^{d_H} dl \, h^2_c(f,l,z)\frac{d^2R}{dz \, dl} \ ,
\end{align}
where $z_F$ and $z_*$ denote the redshifts at the time of string network entering the scaling regime after the horizon reentry of the strings and disappearance, and we have taken the particle horizon $d_H$ at redshift $z$ as the upper limit of integration on $l$. We have taken $z_F$ at a time two orders of magnitude higher than the horizon reentry time of the strings following the velocity-dependent one scale model \cite{Martins:1995tg, Martins:1996jp, Martins:2000cs} (for a recent discussion see \cite{Gouttenoire:2019kij}). It is worth mentioning that the strings reenter the horizon in a sufficiently late universe much higher than the time scale of friction domination, $t_{\rm Pl}/(G\mu)^2$ \cite{Vilenkin:1991yd, Garriga:1993gj} ($t_{\rm Pl}$ is the Planck time).

Fig.~\ref{fig:gws} depicts the stochastic gravitational wave backgrounds sourced by the scalar perturbations and metastable strings. In the case of BP1, the metastable strings provide the dominant contribution to the gravitational wave background and can explain the NANOGrav 15 year data. Moreover, there will be a significant PBH dark matter abundance which can be observed in future lensing data (see Fig.~\ref{fig:fPBH}). On the other hand, if the strings experience about 40 $e$-foldings of inflation and the PTA data can be explained around the nHz frequencies from a combination of the two sources. Moreover, the gravitational wave background displays a UV tail varying as $\Omega_{\rm GW} \propto f^{-1/3}$ starting from the $\mu$Hz frequencies which can be probed in several planned experiments such as $\mu$Ares \cite{Sesana:2019vho}, LISA \cite{Bartolo:2016ami, amaroseoane2017laser}, DECIGO \cite{Sato_2017}, BBO \cite{Crowder:2005nr, Corbin:2005ny}, CE \cite{Regimbau:2016ike}, and ET \cite{Mentasti:2020yyd}. The $f^{-1/3}$ UV tail starts around the mHz frequency in the former case, and the background can be detected in the proposed and ongoing experiments including HLVK \cite{KAGRA:2013rdx}. 
%
\section{Primordial black holes}
\label{sec:pbh}
\begin{figure}[htbp!]
    \centering
    \includegraphics[width=0.8\linewidth]{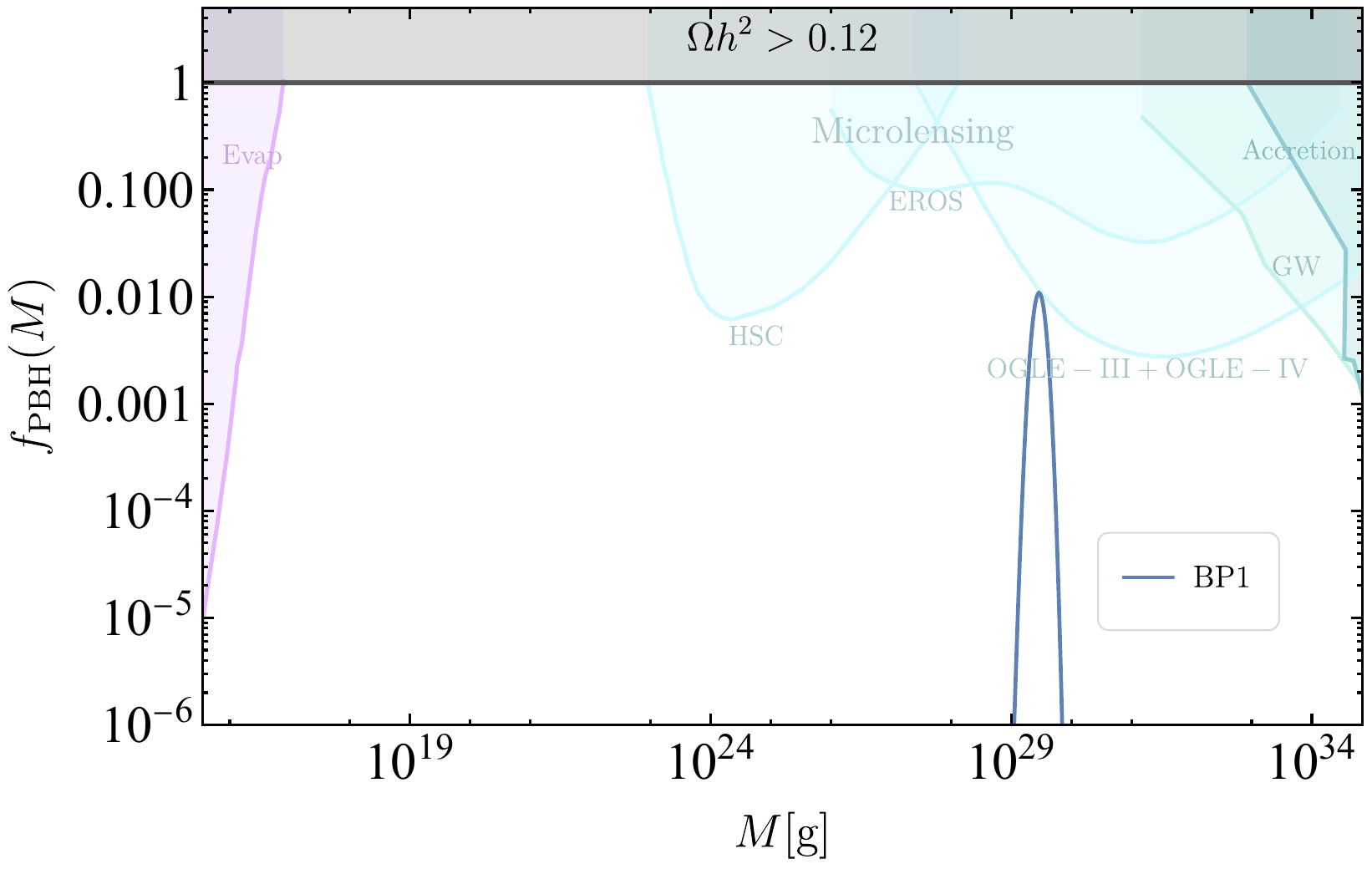}
    \caption{\label{fig:fPBH} Primordial black holes  fractional abundance for BP1. PBHs generated from BP2 have negligibly small abundance.}
\end{figure}
The enhancement in the power spectrum may result in the formation of PBHs if the density contrast $\delta \equiv \delta \rho / \rho$ exceeds a critical value $\delta_c(k)$~\cite{Young:2014ana}. In this case, the PBHs mass fraction $\beta$ compared to the total mass of the universe can be evaluated using the Press-Schechter formalism, namely,
\bea
\label{eq:beta}
\beta(k)= \frac{1}{\sqrt{2 \pi \sigma ^2 (k)}} \int^{\infty}_{\delta_c(k)} d\delta \,  \exp \left(  -\frac{\delta ^2}{2 \sigma^2(k) } \right) \, ,
\eea
where $\sigma$ is the variance of the curvature perturbations which can be computed using the power spectrum and  a window function with Gaussian distribution function $ \widetilde{W}(x)=e^{-x^2/2} $~\cite{Harada:2013epa,Heurtier:2022rhf},
\begin{equation}
\label{eq:sigma2}
\sigma^2 \left(k  \right)= \frac{16}{81}  \int \frac{dk' }{k'} \left(\frac{k'}{k}\right)^4 P_{\zeta}(k') \widetilde{W}\left(\frac{k'}{k}\right),
\end{equation}
with $0.4 \lesssim \delta_c \lesssim 0.6$~\cite{Harada:2013epa,Musco:2004ak,Musco:2008hv,Musco:2012au,Musco:2018rwt,Escriva:2019phb,Escriva:2020tak,Musco:2020jjb,Ghoshal:2023wri,Ijaz:2024zma}. In terms of the co-moving wave number $k$, the PBH mass (in grams) is given by \cite{Ballesteros:2017fsr}
\begin{equation}
\label{eq:MPBHk}
M_{\mathrm{PBH}}(k)=10^{18} \left( \frac{\gamma}{0.2} \right) \left(\frac{g_*(T_f)}{106.75}\right)^{-1/6} \left(\frac{k}{7 \times 10^{13} \, \mathrm{Mpc}^{-1}  }\right)^{-2}  ,
\end{equation}
where we assume that PBHs are created during the radiation epoch, and $T_f$  represents the temperature  at the time of their formation. The fractional abundance of PBHs, $f_{\rm PBH}\equiv \Omega_{\rm PBH} / \Omega_{\rm DM}$ has the form
\begin{equation}
\label{eq:fPBH}
f_{\rm PBH}(M_{\mathrm{PBH}})= 
\frac{\beta(M_{\mathrm{PBH}})}{8 \times 10^{-16}} \left(\frac{\gamma}{0.2}\right)^{3/2} 
       \left(\frac{g_*(T_f)}{106.75}\right)^{-1/4}\left(\frac{M_{\mathrm{PBH}} \,  }{10^{-18 }\; \mathrm{ grams} }\right)^{-1/2}\, , 
\end{equation}
where the factor $\gamma\sim 0.2$ represents the dependence on the gravitational collapse~\cite{Carr:1975qj}. 

Figure~\ref{fig:fPBH} shows the predicted dark matter abundance of PBHs with mass $M=2.9\times 10^{29}$ g ($1.46\times 10^{-4} \, M_\odot$) for BP1, that
constitutes about 1\% of the observed dark matter density which can be detected in the next generation of lensing experiments. For BP2, where the PTA results are explained in terms of SIGW and MSSGW, the corresponding PBH fractional abundance is extremely small. The shaded regions represent the various observational constraints on PBHs from black hole evaporation, accretion and GWs~\cite{Laha:2019ssq,Dasgupta:2019cae,Carr:2020gox,Green:2020jor,Laha:2020ivk,Ray:2021mxu,Saha:2021pqf,Alexandre:2024nuo,Thoss:2024hsr,Dvali:2020wft,Dvali:2018ytn,Hamaide:2023ayu,Virbhadra:1999nm}, and microlensing, including HSC, EROS and OGLE experiments~\cite{Mroz:2024mse}. 
%
\section{Conclusions}
\label{sec:conc}
We have explored the cosmological implications of a realistic flipped SU(5) model of hybrid inflation that yields superheavy metastable cosmic strings during the waterfall phase transition. We show how two sources of gravitational waves  (GWs) appear, namely from the metastable cosmic strings (MSSGW), as well as from scalar perturbations associated with the waterfall field, often referred to as scalar induced gravitational waves (SIGW). We display representative benchmark points that allow for both types of GWs compatible with the recent PTA/NANOGrav results, and also with the LIGO-VIRGO third run. The spectrum features a $f^{-1/3}$ UV tail that can be detected in the proposed and ongoing experiments including HLVK. We also show that PBHs can be produced with mass of around $10^{29}$ g for the case where the MSSGW spectrum is the dominant one. Despite the relatively small dark matter fractional abundance of PBHs, their presence can be tested in the future lensing experiments. 
%
\acknowledgments
R.M. is supported by Institute for Basic Science under the project code: IBS-R018-D3. R.M. and Q.S. would like to thank Professor Masahide Yamaguchi and his colleagues, students and staff for the hospitality provided at the IBS-CTPU-CGA, Tokyo Tech, USTC 2024 Summer Workshop and School on Cosmology, Gravity and Particle Physics, which gave us the opportunity to discuss this project in person.

\bibliographystyle{JHEP}
\bibliography{GUT_TD.bib}

\providecommand{\href}[2]{#2}\begingroup\raggedright\begin{thebibliography}{100}

\bibitem{Georgi:1974sy}
H.~Georgi and S.L.~Glashow, \emph{{Unity of All Elementary Particle Forces}},
  \href{https://doi.org/10.1103/PhysRevLett.32.438}{\emph{Phys. Rev. Lett.}
  {\bfseries 32} (1974) 438}.

\bibitem{Fritzsch:1974nn}
H.~Fritzsch and P.~Minkowski, \emph{{Unified Interactions of Leptons and
  Hadrons}}, \href{https://doi.org/10.1016/0003-4916(75)90211-0}{\emph{Annals
  Phys.} {\bfseries 93} (1975) 193}.

\bibitem{Shafi:1978gg}
Q.~Shafi, \emph{{E(6) as a Unifying Gauge Symmetry}},
  \href{https://doi.org/10.1016/0370-2693(78)90248-4}{\emph{Phys. Lett. B}
  {\bfseries 79} (1978) 301}.

\bibitem{tHooft:1974kcl}
G.~'t~Hooft, \emph{{Magnetic Monopoles in Unified Gauge Theories}},
  \href{https://doi.org/10.1016/0550-3213(74)90486-6}{\emph{Nucl. Phys. B}
  {\bfseries 79} (1974) 276}.

\bibitem{Lazarides:1980cc}
G.~Lazarides, M.~Magg and Q.~Shafi, \emph{{Phase Transitions and Magnetic
  Monopoles in SO(10)}},
  \href{https://doi.org/10.1016/0370-2693(80)90553-5}{\emph{Phys. Lett. B}
  {\bfseries 97} (1980) 87}.

\bibitem{Maji:2022jzu}
R.~Maji and Q.~Shafi, \emph{{Monopoles, strings and gravitational waves in
  non-minimal inflation}},
  \href{https://doi.org/10.1088/1475-7516/2023/03/007}{\emph{JCAP} {\bfseries
  03} (2023) 007} [\href{https://arxiv.org/abs/2208.08137}{{\ttfamily
  2208.08137}}].

\bibitem{Moursy:2024hll}
A.~Moursy and Q.~Shafi, \emph{{Primordial monopoles, black holes and
  gravitational waves}},
  \href{https://doi.org/10.1088/1475-7516/2024/08/064}{\emph{JCAP} {\bfseries
  08} (2024) 064} [\href{https://arxiv.org/abs/2405.04397}{{\ttfamily
  2405.04397}}].

\bibitem{Kibble:1982dd}
T.W.B.~Kibble, G.~Lazarides and Q.~Shafi, \emph{{Walls Bounded by Strings}},
  \href{https://doi.org/10.1103/PhysRevD.26.435}{\emph{Phys. Rev. D} {\bfseries
  26} (1982) 435}.

\bibitem{Kibble:1982ae}
T.W.B.~Kibble, G.~Lazarides and Q.~Shafi, \emph{{Strings in SO(10)}},
  \href{https://doi.org/10.1016/0370-2693(82)90829-2}{\emph{Phys. Lett. B}
  {\bfseries 113} (1982) 237}.

\bibitem{Vilenkin:1982ks}
A.~Vilenkin and A.E.~Everett, \emph{{Cosmic Strings and Domain Walls in Models
  with Goldstone and PseudoGoldstone Bosons}},
  \href{https://doi.org/10.1103/PhysRevLett.48.1867}{\emph{Phys. Rev. Lett.}
  {\bfseries 48} (1982) 1867}.

\bibitem{Lazarides:2019xai}
G.~Lazarides and Q.~Shafi, \emph{{Monopoles, Strings, and Necklaces in $SO(10)$
  and $E_6$}}, \href{https://doi.org/10.1007/JHEP10(2019)193}{\emph{JHEP}
  {\bfseries 10} (2019) 193}
  [\href{https://arxiv.org/abs/1904.06880}{{\ttfamily 1904.06880}}].

\bibitem{Lazarides:2023iim}
G.~Lazarides, Q.~Shafi and A.~Tiwari, \emph{{Composite topological structures
  in SO(10)}}, \href{https://doi.org/10.1007/JHEP05(2023)119}{\emph{JHEP}
  {\bfseries 05} (2023) 119}
  [\href{https://arxiv.org/abs/2303.15159}{{\ttfamily 2303.15159}}].

\bibitem{Maji:2024pll}
R.~Maji, Q.~Shafi and A.~Tiwari, \emph{{Topological structures, dark matter and
  gravitational waves in E$_{6}$}},
  \href{https://doi.org/10.1007/JHEP08(2024)060}{\emph{JHEP} {\bfseries 08}
  (2024) 060} [\href{https://arxiv.org/abs/2406.06308}{{\ttfamily
  2406.06308}}].

\bibitem{Ambrosio:2002qq}
{\scshape MACRO} collaboration, \emph{{Final results of magnetic monopole
  searches with the MACRO experiment}},
  \href{https://doi.org/10.1140/epjc/s2002-01046-9}{\emph{Eur. Phys. J. C}
  {\bfseries 25} (2002) 511}
  [\href{https://arxiv.org/abs/hep-ex/0207020}{{\ttfamily hep-ex/0207020}}].

\bibitem{IceCube:2021eye}
{\scshape IceCube} collaboration, \emph{{Search for Relativistic Magnetic
  Monopoles with Eight Years of IceCube Data}},
  \href{https://doi.org/10.1103/PhysRevLett.128.051101}{\emph{Phys. Rev. Lett.}
  {\bfseries 128} (2022) 051101}
  [\href{https://arxiv.org/abs/2109.13719}{{\ttfamily 2109.13719}}].

\bibitem{ANTARES:2022zbr}
{\scshape ANTARES} collaboration, \emph{{Search for magnetic monopoles with ten
  years of the ANTARES neutrino telescope}},
  \href{https://doi.org/10.1016/j.jheap.2022.03.001}{\emph{JHEAp} {\bfseries
  34} (2022) 1} [\href{https://arxiv.org/abs/2202.13786}{{\ttfamily
  2202.13786}}].

\bibitem{Antusch:2023zjk}
S.~Antusch, K.~Hinze, S.~Saad and J.~Steiner, \emph{{Singling out SO(10) GUT
  models using recent PTA results}},
  \href{https://doi.org/10.1103/PhysRevD.108.095053}{\emph{Phys. Rev. D}
  {\bfseries 108} (2023) 095053}
  [\href{https://arxiv.org/abs/2307.04595}{{\ttfamily 2307.04595}}].

\bibitem{Lazarides:2023rqf}
G.~Lazarides, R.~Maji, A.~Moursy and Q.~Shafi, \emph{{Inflation, superheavy
  metastable strings and gravitational waves in non-supersymmetric flipped
  SU(5)}}, \href{https://doi.org/10.1088/1475-7516/2024/03/006}{\emph{JCAP}
  {\bfseries 03} (2024) 006}
  [\href{https://arxiv.org/abs/2308.07094}{{\ttfamily 2308.07094}}].

\bibitem{Maji:2023fhv}
R.~Maji and W.-I.~Park, \emph{{Supersymmetric U(1)B-L flat direction and
  NANOGrav 15 year data}},
  \href{https://doi.org/10.1088/1475-7516/2024/01/015}{\emph{JCAP} {\bfseries
  01} (2024) 015} [\href{https://arxiv.org/abs/2308.11439}{{\ttfamily
  2308.11439}}].

\bibitem{Ahmed:2023rky}
W.~Ahmed, M.U.~Rehman and U.~Zubair, \emph{{Probing stochastic gravitational
  wave background from $SU(5) \times U(1)_{\chi}$ strings in light of NANOGrav
  15-year data}},
  \href{https://doi.org/10.1088/1475-7516/2024/01/049}{\emph{JCAP} {\bfseries
  01} (2024) 049} [\href{https://arxiv.org/abs/2308.09125}{{\ttfamily
  2308.09125}}].

\bibitem{afzal:2023kqs}
A.~Afzal, Q.~Shafi and A.~Tiwari, \emph{{Gravitational wave emission from
  metastable current-carrying strings in E6}},
  \href{https://doi.org/10.1016/j.physletb.2024.138516}{\emph{Phys. Lett. B}
  {\bfseries 850} (2024) 138516}
  [\href{https://arxiv.org/abs/2311.05564}{{\ttfamily 2311.05564}}].

\bibitem{Antusch:2024nqg}
S.~Antusch, K.~Hinze and S.~Saad, \emph{{Explaining PTA Results by Metastable
  Cosmic Strings from SO(10) GUT}},
  \href{https://arxiv.org/abs/2406.17014}{{\ttfamily 2406.17014}}.

\bibitem{Pallis:2024mip}
C.~Pallis, \emph{{PeV-Scale SUSY and Cosmic Strings from F-Term Hybrid
  Inflation}}, \href{https://doi.org/10.3390/universe10050211}{\emph{Universe}
  {\bfseries 10} (2024) } [\href{https://arxiv.org/abs/2403.09385}{{\ttfamily
  2403.09385}}].

\bibitem{Ahmed:2024iyd}
W.~Ahmed, M.~Mehmood, M.U.~Rehman and U.~Zubair, \emph{{Inflation, Proton Decay
  and Gravitational Waves from Metastable Strings in $SU(4)_C \times SU(2)_L
  \times U(1)_R$ Model}},  \href{https://arxiv.org/abs/2404.06008}{{\ttfamily
  2404.06008}}.

\bibitem{NANOGrav:2023gor}
{\scshape NANOGrav} collaboration, \emph{{The NANOGrav 15 yr Data Set: Evidence
  for a Gravitational-wave Background}},
  \href{https://doi.org/10.3847/2041-8213/acdac6}{\emph{Astrophys. J. Lett.}
  {\bfseries 951} (2023) L8}
  [\href{https://arxiv.org/abs/2306.16213}{{\ttfamily 2306.16213}}].

\bibitem{NANOGrav:2023hfp}
{\scshape NANOGrav} collaboration, \emph{{The NANOGrav 15 yr Data Set:
  Constraints on Supermassive Black Hole Binaries from the Gravitational-wave
  Background}},
  \href{https://doi.org/10.3847/2041-8213/ace18b}{\emph{Astrophys. J. Lett.}
  {\bfseries 952} (2023) L37}
  [\href{https://arxiv.org/abs/2306.16220}{{\ttfamily 2306.16220}}].

\bibitem{NANOGrav:2023hvm}
{\scshape NANOGrav} collaboration, \emph{{The NANOGrav 15 yr Data Set: Search
  for Signals from New Physics}},
  \href{https://doi.org/10.3847/2041-8213/acdc91}{\emph{Astrophys. J. Lett.}
  {\bfseries 951} (2023) L11}
  [\href{https://arxiv.org/abs/2306.16219}{{\ttfamily 2306.16219}}] [Erratum:
  Astrophys.J.Lett. 971, L27 (2024)].

\bibitem{EPTA:2023fyk}
{\scshape EPTA, InPTA:} collaboration, \emph{{The second data release from the
  European Pulsar Timing Array - III. Search for gravitational wave signals}},
  \href{https://doi.org/10.1051/0004-6361/202346844}{\emph{Astron. Astrophys.}
  {\bfseries 678} (2023) A50}
  [\href{https://arxiv.org/abs/2306.16214}{{\ttfamily 2306.16214}}].

\bibitem{EPTA:2023xxk}
{\scshape EPTA, InPTA} collaboration, \emph{{The second data release from the
  European Pulsar Timing Array - IV. Implications for massive black holes, dark
  matter, and the early Universe}},
  \href{https://doi.org/10.1051/0004-6361/202347433}{\emph{Astron. Astrophys.}
  {\bfseries 685} (2024) A94}
  [\href{https://arxiv.org/abs/2306.16227}{{\ttfamily 2306.16227}}].

\bibitem{Reardon:2023gzh}
D.J.~Reardon et~al., \emph{{Search for an Isotropic Gravitational-wave
  Background with the Parkes Pulsar Timing Array}},
  \href{https://doi.org/10.3847/2041-8213/acdd02}{\emph{Astrophys. J. Lett.}
  {\bfseries 951} (2023) L6}
  [\href{https://arxiv.org/abs/2306.16215}{{\ttfamily 2306.16215}}].

\bibitem{Xu_2023}
H.~Xu, S.~Chen, Y.~Guo, J.~Jiang, B.~Wang, J.~Xu et~al., \emph{Searching for
  the nano-hertz stochastic gravitational wave background with the chinese
  pulsar timing array data release i},
  \href{https://doi.org/10.1088/1674-4527/acdfa5}{\emph{Research in Astronomy
  and Astrophysics} {\bfseries 23} (2023) 075024}.

\bibitem{LIGOScientific:2021nrg}
{\scshape LIGO Scientific, Virgo, KAGRA} collaboration, \emph{{Constraints on
  Cosmic Strings Using Data from the Third Advanced LIGO\textendash{}Virgo
  Observing Run}},
  \href{https://doi.org/10.1103/PhysRevLett.126.241102}{\emph{Phys. Rev. Lett.}
  {\bfseries 126} (2021) 241102}
  [\href{https://arxiv.org/abs/2101.12248}{{\ttfamily 2101.12248}}].

\bibitem{Datta:2024bqp}
S.~Datta and R.~Samanta, \emph{{Cosmic superstrings, metastable strings and
  ultralight primordial black holes: from NANOGrav to LIGO and beyond}},
  \href{https://arxiv.org/abs/2409.03498}{{\ttfamily 2409.03498}}.

\bibitem{Lazarides:2022jgr}
G.~Lazarides, R.~Maji and Q.~Shafi, \emph{{Gravitational waves from
  quasi-stable strings}},
  \href{https://doi.org/10.1088/1475-7516/2022/08/042}{\emph{JCAP} {\bfseries
  08} (2022) 042} [\href{https://arxiv.org/abs/2203.11204}{{\ttfamily
  2203.11204}}].

\bibitem{Maji:2023fba}
R.~Maji, W.-I.~Park and Q.~Shafi, \emph{{Gravitational waves from walls bounded
  by strings in SO(10) model of pseudo-Goldstone dark matter}},
  \href{https://doi.org/10.1016/j.physletb.2023.138127}{\emph{Phys. Lett. B}
  {\bfseries 845} (2023) 138127}
  [\href{https://arxiv.org/abs/2305.11775}{{\ttfamily 2305.11775}}].

\bibitem{Lazarides:2023ksx}
G.~Lazarides, R.~Maji and Q.~Shafi, \emph{{Superheavy quasistable strings and
  walls bounded by strings in the light of NANOGrav 15~year data}},
  \href{https://doi.org/10.1103/PhysRevD.108.095041}{\emph{Phys. Rev. D}
  {\bfseries 108} (2023) 095041}
  [\href{https://arxiv.org/abs/2306.17788}{{\ttfamily 2306.17788}}].

\bibitem{Maji:2024tzg}
R.~Maji and Q.~Shafi, \emph{{Kinetic mixing, proton decay and gravitational
  waves in SO(10)}}, \href{https://doi.org/10.1007/JHEP10(2024)157}{\emph{JHEP}
  {\bfseries 10} (2024) 157}
  [\href{https://arxiv.org/abs/2408.14350}{{\ttfamily 2408.14350}}].

\bibitem{Garcia-Bellido:1996mdl}
J.~Garcia-Bellido, A.D.~Linde and D.~Wands, \emph{{Density perturbations and
  black hole formation in hybrid inflation}},
  \href{https://doi.org/10.1103/PhysRevD.54.6040}{\emph{Phys. Rev. D}
  {\bfseries 54} (1996) 6040}
  [\href{https://arxiv.org/abs/astro-ph/9605094}{{\ttfamily
  astro-ph/9605094}}].

\bibitem{Clesse:2015wea}
S.~Clesse and J.~Garc\'\i{}a-Bellido, \emph{{Massive Primordial Black Holes
  from Hybrid Inflation as Dark Matter and the seeds of Galaxies}},
  \href{https://doi.org/10.1103/PhysRevD.92.023524}{\emph{Phys. Rev. D}
  {\bfseries 92} (2015) 023524}
  [\href{https://arxiv.org/abs/1501.07565}{{\ttfamily 1501.07565}}].

\bibitem{Braglia:2022phb}
M.~Braglia, A.~Linde, R.~Kallosh and F.~Finelli, \emph{{Hybrid
  \ensuremath{\alpha}-attractors, primordial black holes and gravitational wave
  backgrounds}},
  \href{https://doi.org/10.1088/1475-7516/2023/04/033}{\emph{JCAP} {\bfseries
  04} (2023) 033} [\href{https://arxiv.org/abs/2211.14262}{{\ttfamily
  2211.14262}}].

\bibitem{Ananda:2006af}
K.N.~Ananda, C.~Clarkson and D.~Wands, \emph{{The Cosmological gravitational
  wave background from primordial density perturbations}},
  \href{https://doi.org/10.1103/PhysRevD.75.123518}{\emph{Phys. Rev. D}
  {\bfseries 75} (2007) 123518}
  [\href{https://arxiv.org/abs/gr-qc/0612013}{{\ttfamily gr-qc/0612013}}].

\bibitem{Baumann:2007zm}
D.~Baumann, P.J.~Steinhardt, K.~Takahashi and K.~Ichiki, \emph{{Gravitational
  Wave Spectrum Induced by Primordial Scalar Perturbations}},
  \href{https://doi.org/10.1103/PhysRevD.76.084019}{\emph{Phys. Rev. D}
  {\bfseries 76} (2007) 084019}
  [\href{https://arxiv.org/abs/hep-th/0703290}{{\ttfamily hep-th/0703290}}].

\bibitem{DeLuca:2020agl}
V.~De~Luca, G.~Franciolini and A.~Riotto, \emph{{NANOGrav Data Hints at
  Primordial Black Holes as Dark Matter}},
  \href{https://doi.org/10.1103/PhysRevLett.126.041303}{\emph{Phys. Rev. Lett.}
  {\bfseries 126} (2021) 041303}
  [\href{https://arxiv.org/abs/2009.08268}{{\ttfamily 2009.08268}}].

\bibitem{Escriva:2022duf}
A.~Escriv\`a, F.~Kuhnel and Y.~Tada, \emph{{Primordial Black Holes}},
  \href{https://arxiv.org/abs/2211.05767}{{\ttfamily 2211.05767}}.

\bibitem{Khlopov:2008qy}
M.Y.~Khlopov, \emph{{Primordial Black Holes}},
  \href{https://doi.org/10.1088/1674-4527/10/6/001}{\emph{Res. Astron.
  Astrophys.} {\bfseries 10} (2010) 495}
  [\href{https://arxiv.org/abs/0801.0116}{{\ttfamily 0801.0116}}].

\bibitem{Belotsky:2014kca}
K.M.~Belotsky, A.D.~Dmitriev, E.A.~Esipova, V.A.~Gani, A.V.~Grobov,
  M.Y.~Khlopov et~al., \emph{{Signatures of primordial black hole dark
  matter}}, \href{https://doi.org/10.1142/S0217732314400057}{\emph{Mod. Phys.
  Lett. A} {\bfseries 29} (2014) 1440005}
  [\href{https://arxiv.org/abs/1410.0203}{{\ttfamily 1410.0203}}].

\bibitem{Dealtry:2019ldr}
{\scshape Hyper-Kamiokande} collaboration, \emph{{Hyper-Kamiokande}},  in
  \emph{{Prospects in Neutrino Physics}}, 4, 2019
  [\href{https://arxiv.org/abs/1904.10206}{{\ttfamily 1904.10206}}].

\bibitem{Preskill:1992ck}
J.~Preskill and A.~Vilenkin, \emph{{Decay of metastable topological defects}},
  \href{https://doi.org/10.1103/PhysRevD.47.2324}{\emph{Phys. Rev. D}
  {\bfseries 47} (1993) 2324}
  [\href{https://arxiv.org/abs/hep-ph/9209210}{{\ttfamily hep-ph/9209210}}].

\bibitem{Ibrahim:2022cqs}
M.~Ibrahim, M.~Ashry, E.~Elkhateeb, A.M.~Awad and A.~Moursy, \emph{{Modified
  hybrid inflation, reheating, and stabilization of the electroweak vacuum}},
  \href{https://doi.org/10.1103/PhysRevD.107.035023}{\emph{Phys. Rev. D}
  {\bfseries 107} (2023) 035023}
  [\href{https://arxiv.org/abs/2210.03247}{{\ttfamily 2210.03247}}].

\bibitem{Linde:1993cn}
A.D.~Linde, \emph{{Hybrid inflation}},
  \href{https://doi.org/10.1103/PhysRevD.49.748}{\emph{Phys. Rev. D} {\bfseries
  49} (1994) 748} [\href{https://arxiv.org/abs/astro-ph/9307002}{{\ttfamily
  astro-ph/9307002}}].

\bibitem{Clesse:2010iz}
S.~Clesse, \emph{{Hybrid inflation along waterfall trajectories}},
  \href{https://doi.org/10.1103/PhysRevD.83.063518}{\emph{Phys. Rev. D}
  {\bfseries 83} (2011) 063518}
  [\href{https://arxiv.org/abs/1006.4522}{{\ttfamily 1006.4522}}].

\bibitem{Kodama:2011vs}
H.~Kodama, K.~Kohri and K.~Nakayama, \emph{{On the waterfall behavior in hybrid
  inflation}}, \href{https://doi.org/10.1143/PTP.126.331}{\emph{Prog. Theor.
  Phys.} {\bfseries 126} (2011) 331}
  [\href{https://arxiv.org/abs/1102.5612}{{\ttfamily 1102.5612}}].

\bibitem{Ringeval:2007am}
C.~Ringeval, \emph{{The exact numerical treatment of inflationary models}},
  \href{https://doi.org/10.1007/978-3-540-74353-8_7}{\emph{Lect. Notes Phys.}
  {\bfseries 738} (2008) 243}
  [\href{https://arxiv.org/abs/astro-ph/0703486}{{\ttfamily
  astro-ph/0703486}}].

\bibitem{Clesse:2013jra}
S.~Clesse, B.~Garbrecht and Y.~Zhu, \emph{{Non-Gaussianities and Curvature
  Perturbations from Hybrid Inflation}},
  \href{https://doi.org/10.1103/PhysRevD.89.063519}{\emph{Phys. Rev. D}
  {\bfseries 89} (2014) 063519}
  [\href{https://arxiv.org/abs/1304.7042}{{\ttfamily 1304.7042}}].

\bibitem{BICEP:2021xfz}
{\scshape BICEP, Keck} collaboration, \emph{{Improved Constraints on Primordial
  Gravitational Waves using Planck, WMAP, and BICEP/Keck Observations through
  the 2018 Observing Season}},
  \href{https://doi.org/10.1103/PhysRevLett.127.151301}{\emph{Phys. Rev. Lett.}
  {\bfseries 127} (2021) 151301}
  [\href{https://arxiv.org/abs/2110.00483}{{\ttfamily 2110.00483}}].

\bibitem{Planck:2018jri}
{\scshape Planck} collaboration, \emph{{Planck 2018 results. X. Constraints on
  inflation}}, \href{https://doi.org/10.1051/0004-6361/201833887}{\emph{Astron.
  Astrophys.} {\bfseries 641} (2020) A10}
  [\href{https://arxiv.org/abs/1807.06211}{{\ttfamily 1807.06211}}].

\bibitem{Liddle:2003as}
A.R.~Liddle and S.M.~Leach, \emph{{How long before the end of inflation were
  observable perturbations produced?}},
  \href{https://doi.org/10.1103/PhysRevD.68.103503}{\emph{Phys. Rev. D}
  {\bfseries 68} (2003) 103503}
  [\href{https://arxiv.org/abs/astro-ph/0305263}{{\ttfamily
  astro-ph/0305263}}].

\bibitem{Chakrabortty:2020otp}
J.~Chakrabortty, G.~Lazarides, R.~Maji and Q.~Shafi, \emph{{Primordial
  Monopoles and Strings, Inflation, and Gravity Waves}},
  \href{https://doi.org/10.1007/JHEP02(2021)114}{\emph{JHEP} {\bfseries 02}
  (2021) 114} [\href{https://arxiv.org/abs/2011.01838}{{\ttfamily
  2011.01838}}].

\bibitem{Kawai:2023dac}
S.~Kawai and N.~Okada, \emph{{Reheating consistency condition on the
  classically conformal U(1)B\textendash{}L Higgs inflation model}},
  \href{https://doi.org/10.1103/PhysRevD.108.015013}{\emph{Phys. Rev. D}
  {\bfseries 108} (2023) 015013}
  [\href{https://arxiv.org/abs/2303.00342}{{\ttfamily 2303.00342}}].

\bibitem{Senoguz:2015lba}
V.N.~\c{S}eno\u{g}uz and Q.~Shafi, \emph{{Primordial monopoles, proton decay,
  gravity waves and GUT inflation}},
  \href{https://doi.org/10.1016/j.physletb.2015.11.037}{\emph{Phys. Lett. B}
  {\bfseries 752} (2016) 169}
  [\href{https://arxiv.org/abs/1510.04442}{{\ttfamily 1510.04442}}].

\bibitem{Maity:2024odg}
S.~Maity, N.~Bhaumik, M.R.~Haque, D.~Maity and L.~Sriramkumar,
  \emph{{Constraining the history of reheating with the NANOGrav 15-year
  data}},  \href{https://arxiv.org/abs/2403.16963}{{\ttfamily 2403.16963}}.

\bibitem{Lazarides:1997xr}
G.~Lazarides, \emph{{Inflation}},  in \emph{{6th BCSPIN Kathmandu Summer School
  in Physics: Current Trends in High-Energy Physics and Cosmology}}, 5, 1997
  [\href{https://arxiv.org/abs/hep-ph/9802415}{{\ttfamily hep-ph/9802415}}].

\bibitem{Lazarides:2001zd}
G.~Lazarides, \emph{{Inflationary cosmology}},
  \href{https://doi.org/10.1007/3-540-48025-0_13}{\emph{Lect. Notes Phys.}
  {\bfseries 592} (2002) 351}
  [\href{https://arxiv.org/abs/hep-ph/0111328}{{\ttfamily hep-ph/0111328}}].

\bibitem{Nakama:2014vla}
T.~Nakama, T.~Suyama and J.~Yokoyama, \emph{{Reheating the Universe Once More:
  The Dissipation of Acoustic Waves as a Novel Probe of Primordial
  Inhomogeneities on Even Smaller Scales}},
  \href{https://doi.org/10.1103/PhysRevLett.113.061302}{\emph{Phys. Rev. Lett.}
  {\bfseries 113} (2014) 061302}
  [\href{https://arxiv.org/abs/1403.5407}{{\ttfamily 1403.5407}}].

\bibitem{Fixsen:1996nj}
D.J.~Fixsen, E.S.~Cheng, J.M.~Gales, J.C.~Mather, R.A.~Shafer and E.L.~Wright,
  \emph{{The Cosmic Microwave Background spectrum from the full COBE FIRAS data
  set}}, \href{https://doi.org/10.1086/178173}{\emph{Astrophys. J.} {\bfseries
  473} (1996) 576} [\href{https://arxiv.org/abs/astro-ph/9605054}{{\ttfamily
  astro-ph/9605054}}].

\bibitem{Kogut:2011xw}
A.~Kogut et~al., \emph{{The Primordial Inflation Explorer (PIXIE): A Nulling
  Polarimeter for Cosmic Microwave Background Observations}},
  \href{https://doi.org/10.1088/1475-7516/2011/07/025}{\emph{JCAP} {\bfseries
  07} (2011) 025} [\href{https://arxiv.org/abs/1105.2044}{{\ttfamily
  1105.2044}}].

\bibitem{Green:2020jor}
A.M.~Green and B.J.~Kavanagh, \emph{{Primordial Black Holes as a dark matter
  candidate}}, \href{https://doi.org/10.1088/1361-6471/abc534}{\emph{J. Phys.
  G} {\bfseries 48} (2021) 043001}
  [\href{https://arxiv.org/abs/2007.10722}{{\ttfamily 2007.10722}}].

\bibitem{Lewicki:2021xku}
M.~Lewicki, O.~Pujol\`as and V.~Vaskonen, \emph{{Escape from supercooling with
  or without bubbles: gravitational wave signatures}},
  \href{https://doi.org/10.1140/epjc/s10052-021-09669-6}{\emph{Eur. Phys. J. C}
  {\bfseries 81} (2021) 857}
  [\href{https://arxiv.org/abs/2106.09706}{{\ttfamily 2106.09706}}].

\bibitem{Kohri:2018awv}
K.~Kohri and T.~Terada, \emph{{Semianalytic calculation of gravitational wave
  spectrum nonlinearly induced from primordial curvature perturbations}},
  \href{https://doi.org/10.1103/PhysRevD.97.123532}{\emph{Phys. Rev. D}
  {\bfseries 97} (2018) 123532}
  [\href{https://arxiv.org/abs/1804.08577}{{\ttfamily 1804.08577}}].

\bibitem{Chatterjee:2017hru}
A.~Chatterjee and A.~Mazumdar, \emph{{Observable tensor-to-scalar ratio and
  secondary gravitational wave background}},
  \href{https://doi.org/10.1103/PhysRevD.97.063517}{\emph{Phys. Rev. D}
  {\bfseries 97} (2018) 063517}
  [\href{https://arxiv.org/abs/1708.07293}{{\ttfamily 1708.07293}}].

\bibitem{Espinosa:2018eve}
J.R.~Espinosa, D.~Racco and A.~Riotto, \emph{{A Cosmological Signature of the
  SM Higgs Instability: Gravitational Waves}},
  \href{https://doi.org/10.1088/1475-7516/2018/09/012}{\emph{JCAP} {\bfseries
  09} (2018) 012} [\href{https://arxiv.org/abs/1804.07732}{{\ttfamily
  1804.07732}}].

\bibitem{Inomata:2019yww}
K.~Inomata and T.~Terada, \emph{{Gauge Independence of Induced Gravitational
  Waves}}, \href{https://doi.org/10.1103/PhysRevD.101.023523}{\emph{Phys. Rev.
  D} {\bfseries 101} (2020) 023523}
  [\href{https://arxiv.org/abs/1912.00785}{{\ttfamily 1912.00785}}].

\bibitem{Domenech:2019quo}
G.~Dom\`enech, \emph{{Induced gravitational waves in a general cosmological
  background}}, \href{https://doi.org/10.1142/S0218271820500285}{\emph{Int. J.
  Mod. Phys. D} {\bfseries 29} (2020) 2050028}
  [\href{https://arxiv.org/abs/1912.05583}{{\ttfamily 1912.05583}}].

\bibitem{Domenech:2020kqm}
G.~Dom\`enech, S.~Pi and M.~Sasaki, \emph{{Induced gravitational waves as a
  probe of thermal history of the universe}},
  \href{https://doi.org/10.1088/1475-7516/2020/08/017}{\emph{JCAP} {\bfseries
  08} (2020) 017} [\href{https://arxiv.org/abs/2005.12314}{{\ttfamily
  2005.12314}}].

\bibitem{Domenech:2021ztg}
G.~Dom\`enech, \emph{{Scalar Induced Gravitational Waves Review}},
  \href{https://doi.org/10.3390/universe7110398}{\emph{Universe} {\bfseries 7}
  (2021) 398} [\href{https://arxiv.org/abs/2109.01398}{{\ttfamily
  2109.01398}}].

\bibitem{Basilakos:2023jvp}
S.~Basilakos, D.V.~Nanopoulos, T.~Papanikolaou, E.N.~Saridakis and C.~Tzerefos,
  \emph{{Induced gravitational waves from flipped SU(5) superstring theory at
  nHz}}, \href{https://doi.org/10.1016/j.physletb.2024.138446}{\emph{Phys.
  Lett. B} {\bfseries 849} (2024) 138446}
  [\href{https://arxiv.org/abs/2309.15820}{{\ttfamily 2309.15820}}].

\bibitem{Domenech:2024rks}
G.~Dom\`enech, S.~Pi, A.~Wang and J.~Wang, \emph{{Induced gravitational wave
  interpretation of PTA data: a complete study for general equation of state}},
  \href{https://doi.org/10.1088/1475-7516/2024/08/054}{\emph{JCAP} {\bfseries
  08} (2024) 054} [\href{https://arxiv.org/abs/2402.18965}{{\ttfamily
  2402.18965}}].

\bibitem{Wang:2019zhj}
B.~Wang and Y.~Zhang, \emph{{Second-order cosmological perturbations IV.
  Produced by scalar-tensor and tensor-tensor couplings during the radiation
  dominated stage}},
  \href{https://doi.org/10.1103/PhysRevD.99.123008}{\emph{Phys. Rev. D}
  {\bfseries 99} (2019) 123008}
  [\href{https://arxiv.org/abs/1905.03272}{{\ttfamily 1905.03272}}].

\bibitem{Vachaspati:1984gt}
T.~Vachaspati and A.~Vilenkin, \emph{{Gravitational Radiation from Cosmic
  Strings}}, \href{https://doi.org/10.1103/PhysRevD.31.3052}{\emph{Phys. Rev.
  D} {\bfseries 31} (1985) 3052}.

\bibitem{Kibble:1984hp}
T.W.B.~Kibble, \emph{{Evolution of a system of cosmic strings}},
  \href{https://doi.org/10.1016/0550-3213(85)90596-6}{\emph{Nucl. Phys. B}
  {\bfseries 252} (1985) 227} [Erratum: Nucl.Phys.B 261, 750 (1985)].

\bibitem{Vilenkin:2000jqa}
A.~Vilenkin and E.P.S.~Shellard, \emph{{Cosmic Strings and Other Topological
  Defects}}, Cambridge University Press (7, 2000).

\bibitem{Damour:2001bk}
T.~Damour and A.~Vilenkin, \emph{{Gravitational wave bursts from cusps and
  kinks on cosmic strings}},
  \href{https://doi.org/10.1103/PhysRevD.64.064008}{\emph{Phys. Rev. D}
  {\bfseries 64} (2001) 064008}
  [\href{https://arxiv.org/abs/gr-qc/0104026}{{\ttfamily gr-qc/0104026}}].

\bibitem{Vanchurin:2005pa}
V.~Vanchurin, K.D.~Olum and A.~Vilenkin, \emph{{Scaling of cosmic string
  loops}}, \href{https://doi.org/10.1103/PhysRevD.74.063527}{\emph{Phys. Rev.
  D} {\bfseries 74} (2006) 063527}
  [\href{https://arxiv.org/abs/gr-qc/0511159}{{\ttfamily gr-qc/0511159}}].

\bibitem{Ringeval:2005kr}
C.~Ringeval, M.~Sakellariadou and F.~Bouchet, \emph{{Cosmological evolution of
  cosmic string loops}},
  \href{https://doi.org/10.1088/1475-7516/2007/02/023}{\emph{JCAP} {\bfseries
  02} (2007) 023} [\href{https://arxiv.org/abs/astro-ph/0511646}{{\ttfamily
  astro-ph/0511646}}].

\bibitem{Olum:2006ix}
K.D.~Olum and V.~Vanchurin, \emph{{Cosmic string loops in the expanding
  Universe}}, \href{https://doi.org/10.1103/PhysRevD.75.063521}{\emph{Phys.
  Rev. D} {\bfseries 75} (2007) 063521}
  [\href{https://arxiv.org/abs/astro-ph/0610419}{{\ttfamily
  astro-ph/0610419}}].

\bibitem{Leblond:2009fq}
L.~Leblond, B.~Shlaer and X.~Siemens, \emph{{Gravitational Waves from Broken
  Cosmic Strings: The Bursts and the Beads}},
  \href{https://doi.org/10.1103/PhysRevD.79.123519}{\emph{Phys. Rev. D}
  {\bfseries 79} (2009) 123519}
  [\href{https://arxiv.org/abs/0903.4686}{{\ttfamily 0903.4686}}].

\bibitem{Olmez:2010bi}
S.~Olmez, V.~Mandic and X.~Siemens, \emph{{Gravitational-Wave Stochastic
  Background from Kinks and Cusps on Cosmic Strings}},
  \href{https://doi.org/10.1103/PhysRevD.81.104028}{\emph{Phys. Rev. D}
  {\bfseries 81} (2010) 104028}
  [\href{https://arxiv.org/abs/1004.0890}{{\ttfamily 1004.0890}}].

\bibitem{Blanco-Pillado:2013qja}
J.J.~Blanco-Pillado, K.D.~Olum and B.~Shlaer, \emph{{The number of cosmic
  string loops}}, \href{https://doi.org/10.1103/PhysRevD.89.023512}{\emph{Phys.
  Rev. D} {\bfseries 89} (2014) 023512}
  [\href{https://arxiv.org/abs/1309.6637}{{\ttfamily 1309.6637}}].

\bibitem{Blanco-Pillado:2017oxo}
J.J.~Blanco-Pillado and K.D.~Olum, \emph{{Stochastic gravitational wave
  background from smoothed cosmic string loops}},
  \href{https://doi.org/10.1103/PhysRevD.96.104046}{\emph{Phys. Rev. D}
  {\bfseries 96} (2017) 104046}
  [\href{https://arxiv.org/abs/1709.02693}{{\ttfamily 1709.02693}}].

\bibitem{Cui:2018rwi}
Y.~Cui, M.~Lewicki, D.E.~Morrissey and J.D.~Wells, \emph{{Probing the pre-BBN
  universe with gravitational waves from cosmic strings}},
  \href{https://doi.org/10.1007/JHEP01(2019)081}{\emph{JHEP} {\bfseries 01}
  (2019) 081} [\href{https://arxiv.org/abs/1808.08968}{{\ttfamily
  1808.08968}}].

\bibitem{Buchmuller:2019gfy}
W.~Buchmuller, V.~Domcke, H.~Murayama and K.~Schmitz, \emph{{Probing the scale
  of grand unification with gravitational waves}},
  \href{https://doi.org/10.1016/j.physletb.2020.135764}{\emph{Phys. Lett. B}
  {\bfseries 809} (2020) 135764}
  [\href{https://arxiv.org/abs/1912.03695}{{\ttfamily 1912.03695}}].

\bibitem{Buchmuller:2021mbb}
W.~Buchmuller, V.~Domcke and K.~Schmitz, \emph{{Stochastic gravitational-wave
  background from metastable cosmic strings}},
  \href{https://doi.org/10.1088/1475-7516/2021/12/006}{\emph{JCAP} {\bfseries
  12} (2021) 006} [\href{https://arxiv.org/abs/2107.04578}{{\ttfamily
  2107.04578}}].

\bibitem{Dunsky:2021tih}
D.I.~Dunsky, A.~Ghoshal, H.~Murayama, Y.~Sakakihara and G.~White, \emph{{GUTs,
  hybrid topological defects, and gravitational waves}},
  \href{https://doi.org/10.1103/PhysRevD.106.075030}{\emph{Phys. Rev. D}
  {\bfseries 106} (2022) 075030}
  [\href{https://arxiv.org/abs/2111.08750}{{\ttfamily 2111.08750}}].

\bibitem{Roshan:2024qnv}
R.~Roshan and G.~White, \emph{{Using gravitational waves to see the first
  second of the Universe}},  \href{https://arxiv.org/abs/2401.04388}{{\ttfamily
  2401.04388}}.

\bibitem{Mangano:2011ar}
G.~Mangano and P.D.~Serpico, \emph{{A robust upper limit on $N_{\rm eff}$ from
  BBN, circa 2011}},
  \href{https://doi.org/10.1016/j.physletb.2011.05.075}{\emph{Phys. Lett. B}
  {\bfseries 701} (2011) 296}
  [\href{https://arxiv.org/abs/1103.1261}{{\ttfamily 1103.1261}}].

\bibitem{Thrane:2013oya}
E.~Thrane and J.D.~Romano, \emph{{Sensitivity curves for searches for
  gravitational-wave backgrounds}},
  \href{https://doi.org/10.1103/PhysRevD.88.124032}{\emph{Phys. Rev. D}
  {\bfseries 88} (2013) 124032}
  [\href{https://arxiv.org/abs/1310.5300}{{\ttfamily 1310.5300}}].

\bibitem{Schmitz:2020syl}
K.~Schmitz, \emph{{New Sensitivity Curves for Gravitational-Wave Signals from
  Cosmological Phase Transitions}},
  \href{https://doi.org/10.1007/JHEP01(2021)097}{\emph{JHEP} {\bfseries 01}
  (2021) 097} [\href{https://arxiv.org/abs/2002.04615}{{\ttfamily
  2002.04615}}].

\bibitem{KAGRA:2013rdx}
{\scshape KAGRA, LIGO Scientific, Virgo, VIRGO} collaboration, \emph{{Prospects
  for observing and localizing gravitational-wave transients with Advanced
  LIGO, Advanced Virgo and KAGRA}},
  \href{https://doi.org/10.1007/s41114-020-00026-9}{\emph{Living Rev. Rel.}
  {\bfseries 21} (2018) 3} [\href{https://arxiv.org/abs/1304.0670}{{\ttfamily
  1304.0670}}].

\bibitem{Regimbau:2016ike}
T.~Regimbau, M.~Evans, N.~Christensen, E.~Katsavounidis, B.~Sathyaprakash and
  S.~Vitale, \emph{{Digging deeper: Observing primordial gravitational waves
  below the binary black hole produced stochastic background}},
  \href{https://doi.org/10.1103/PhysRevLett.118.151105}{\emph{Phys. Rev. Lett.}
  {\bfseries 118} (2017) 151105}
  [\href{https://arxiv.org/abs/1611.08943}{{\ttfamily 1611.08943}}].

\bibitem{Mentasti:2020yyd}
G.~Mentasti and M.~Peloso, \emph{{ET sensitivity to the anisotropic Stochastic
  Gravitational Wave Background}},
  \href{https://doi.org/10.1088/1475-7516/2021/03/080}{\emph{JCAP} {\bfseries
  03} (2021) 080} [\href{https://arxiv.org/abs/2010.00486}{{\ttfamily
  2010.00486}}].

\bibitem{Sato_2017}
S.~Sato et~al., \emph{The status of {DECIGO}},
  \href{https://doi.org/10.1088/1742-6596/840/1/012010}{\emph{Journal of
  Physics: Conference Series} {\bfseries 840} (2017) 012010}.

\bibitem{Crowder:2005nr}
J.~Crowder and N.J.~Cornish, \emph{{Beyond LISA: Exploring future gravitational
  wave missions}},
  \href{https://doi.org/10.1103/PhysRevD.72.083005}{\emph{Phys. Rev. D}
  {\bfseries 72} (2005) 083005}
  [\href{https://arxiv.org/abs/gr-qc/0506015}{{\ttfamily gr-qc/0506015}}].

\bibitem{Corbin:2005ny}
V.~Corbin and N.J.~Cornish, \emph{{Detecting the cosmic gravitational wave
  background with the big bang observer}},
  \href{https://doi.org/10.1088/0264-9381/23/7/014}{\emph{Class. Quant. Grav.}
  {\bfseries 23} (2006) 2435}
  [\href{https://arxiv.org/abs/gr-qc/0512039}{{\ttfamily gr-qc/0512039}}].

\bibitem{Bartolo:2016ami}
N.~Bartolo et~al., \emph{{Science with the space-based interferometer LISA. IV:
  Probing inflation with gravitational waves}},
  \href{https://doi.org/10.1088/1475-7516/2016/12/026}{\emph{JCAP} {\bfseries
  12} (2016) 026} [\href{https://arxiv.org/abs/1610.06481}{{\ttfamily
  1610.06481}}].

\bibitem{amaroseoane2017laser}
P.~Amaro-Seoane et~al., \emph{Laser interferometer space antenna},
  \href{https://arxiv.org/abs/1702.00786}{{\ttfamily 1702.00786}}.

\bibitem{5136190}
P.E.~{Dewdney}, P.J.~{Hall}, R.T.~{Schilizzi} and T.J.L.W.~{Lazio}, \emph{The
  square kilometre array},
  \href{https://doi.org/10.1109/JPROC.2009.2021005}{\emph{Proceedings of the
  IEEE} {\bfseries 97} (2009) 1482}.

\bibitem{Janssen:2014dka}
G.~Janssen et~al., \emph{{Gravitational wave astronomy with the SKA}},
  \href{https://doi.org/10.22323/1.215.0037}{\emph{PoS} {\bfseries AASKA14}
  (2015) 037} [\href{https://arxiv.org/abs/1501.00127}{{\ttfamily
  1501.00127}}].

\bibitem{Martins:1995tg}
C.~Martins and E.~Shellard, \emph{Quantitative string evolution},
  \href{https://doi.org/10.1103/PhysRevD.54.2535}{\emph{Phys. Rev. D}
  {\bfseries 54} (1996) 2535}
  [\href{https://arxiv.org/abs/hep-ph/9602271}{{\ttfamily hep-ph/9602271}}].

\bibitem{Martins:1996jp}
C.~Martins and E.~Shellard, \emph{Extending the velocity-dependent one-scale
  string evolution model},
  \href{https://doi.org/10.1103/PhysRevD.53.R575}{\emph{Phys. Rev. D}
  {\bfseries 53} (1996) R575}
  [\href{https://arxiv.org/abs/hep-ph/9603271}{{\ttfamily hep-ph/9603271}}].

\bibitem{Martins:2000cs}
C.~Martins and E.~Shellard, \emph{Scaling of cosmological string networks},
  \href{https://doi.org/10.1103/PhysRevD.65.043514}{\emph{Phys. Rev. D}
  {\bfseries 65} (2002) 043514}
  [\href{https://arxiv.org/abs/hep-ph/0003298}{{\ttfamily hep-ph/0003298}}].

\bibitem{Gouttenoire:2019kij}
Y.~Gouttenoire, G.~Servant and P.~Simakachorn, \emph{{Beyond the Standard
  Models with Cosmic Strings}},
  \href{https://doi.org/10.1088/1475-7516/2020/07/032}{\emph{JCAP} {\bfseries
  07} (2020) 032} [\href{https://arxiv.org/abs/1912.02569}{{\ttfamily
  1912.02569}}].

\bibitem{Vilenkin:1991yd}
A.~Vilenkin, \emph{Cosmic string dynamics with friction},
  \href{https://doi.org/10.1103/PhysRevD.43.1060}{\emph{Phys. Rev. D}
  {\bfseries 43} (1991) 1060}.

\bibitem{Garriga:1993gj}
J.~Garriga and M.~Sakellariadou, \emph{Effects of friction on cosmic strings},
  \href{https://doi.org/10.1103/PhysRevD.48.2502}{\emph{Phys. Rev. D}
  {\bfseries 48} (1993) 2502}
  [\href{https://arxiv.org/abs/hep-th/9303024}{{\ttfamily hep-th/9303024}}].

\bibitem{Sesana:2019vho}
A.~Sesana et~al., \emph{{Unveiling the gravitational universe at $\mu$-Hz
  frequencies}}, \href{https://doi.org/10.1007/s10686-021-09709-9}{\emph{Exper.
  Astron.} {\bfseries 51} (2021) 1333}
  [\href{https://arxiv.org/abs/1908.11391}{{\ttfamily 1908.11391}}].

\bibitem{Young:2014ana}
S.~Young, C.T.~Byrnes and M.~Sasaki, \emph{{Calculating the mass fraction of
  primordial black holes}},
  \href{https://doi.org/10.1088/1475-7516/2014/07/045}{\emph{JCAP} {\bfseries
  07} (2014) 045} [\href{https://arxiv.org/abs/1405.7023}{{\ttfamily
  1405.7023}}].

\bibitem{Harada:2013epa}
T.~Harada, C.-M.~Yoo and K.~Kohri, \emph{{Threshold of primordial black hole
  formation}}, \href{https://doi.org/10.1103/PhysRevD.88.084051}{\emph{Phys.
  Rev. D} {\bfseries 88} (2013) 084051}
  [\href{https://arxiv.org/abs/1309.4201}{{\ttfamily 1309.4201}}] [Erratum:
  Phys.Rev.D 89, 029903 (2014)].

\bibitem{Heurtier:2022rhf}
L.~Heurtier, A.~Moursy and L.~Wacquez, \emph{{Cosmological imprints of SUSY
  breaking in models of sgoldstinoless non-oscillatory inflation}},
  \href{https://doi.org/10.1088/1475-7516/2023/03/020}{\emph{JCAP} {\bfseries
  03} (2023) 020} [\href{https://arxiv.org/abs/2207.11502}{{\ttfamily
  2207.11502}}].

\bibitem{Musco:2004ak}
I.~Musco, J.C.~Miller and L.~Rezzolla, \emph{{Computations of primordial black
  hole formation}},
  \href{https://doi.org/10.1088/0264-9381/22/7/013}{\emph{Class. Quant. Grav.}
  {\bfseries 22} (2005) 1405}
  [\href{https://arxiv.org/abs/gr-qc/0412063}{{\ttfamily gr-qc/0412063}}].

\bibitem{Musco:2008hv}
I.~Musco, J.C.~Miller and A.G.~Polnarev, \emph{{Primordial black hole formation
  in the radiative era: Investigation of the critical nature of the collapse}},
  \href{https://doi.org/10.1088/0264-9381/26/23/235001}{\emph{Class. Quant.
  Grav.} {\bfseries 26} (2009) 235001}
  [\href{https://arxiv.org/abs/0811.1452}{{\ttfamily 0811.1452}}].

\bibitem{Musco:2012au}
I.~Musco and J.C.~Miller, \emph{{Primordial black hole formation in the early
  universe: critical behaviour and self-similarity}},
  \href{https://doi.org/10.1088/0264-9381/30/14/145009}{\emph{Class. Quant.
  Grav.} {\bfseries 30} (2013) 145009}
  [\href{https://arxiv.org/abs/1201.2379}{{\ttfamily 1201.2379}}].

\bibitem{Musco:2018rwt}
I.~Musco, \emph{{Threshold for primordial black holes: Dependence on the shape
  of the cosmological perturbations}},
  \href{https://doi.org/10.1103/PhysRevD.100.123524}{\emph{Phys. Rev. D}
  {\bfseries 100} (2019) 123524}
  [\href{https://arxiv.org/abs/1809.02127}{{\ttfamily 1809.02127}}].

\bibitem{Escriva:2019phb}
A.~Escriv\`a, C.~Germani and R.K.~Sheth, \emph{{Universal threshold for
  primordial black hole formation}},
  \href{https://doi.org/10.1103/PhysRevD.101.044022}{\emph{Phys. Rev. D}
  {\bfseries 101} (2020) 044022}
  [\href{https://arxiv.org/abs/1907.13311}{{\ttfamily 1907.13311}}].

\bibitem{Escriva:2020tak}
A.~Escriv\`a, C.~Germani and R.K.~Sheth, \emph{{Analytical thresholds for black
  hole formation in general cosmological backgrounds}},
  \href{https://doi.org/10.1088/1475-7516/2021/01/030}{\emph{JCAP} {\bfseries
  01} (2021) 030} [\href{https://arxiv.org/abs/2007.05564}{{\ttfamily
  2007.05564}}].

\bibitem{Musco:2020jjb}
I.~Musco, V.~De~Luca, G.~Franciolini and A.~Riotto, \emph{{Threshold for
  primordial black holes. II. A simple analytic prescription}},
  \href{https://doi.org/10.1103/PhysRevD.103.063538}{\emph{Phys. Rev. D}
  {\bfseries 103} (2021) 063538}
  [\href{https://arxiv.org/abs/2011.03014}{{\ttfamily 2011.03014}}].

\bibitem{Ghoshal:2023wri}
A.~Ghoshal, A.~Moursy and Q.~Shafi, \emph{{Cosmological probes of grand
  unification: Primordial black holes and scalar-induced gravitational waves}},
  \href{https://doi.org/10.1103/PhysRevD.108.055039}{\emph{Phys. Rev. D}
  {\bfseries 108} (2023) 055039}
  [\href{https://arxiv.org/abs/2306.04002}{{\ttfamily 2306.04002}}].

\bibitem{Ijaz:2024zma}
N.~Ijaz and M.U.~Rehman, \emph{{Exploring Primordial Black Holes and
  Gravitational Waves with R-Symmetric GUT Higgs Inflation}},
  \href{https://arxiv.org/abs/2402.13924}{{\ttfamily 2402.13924}}.

\bibitem{Ballesteros:2017fsr}
G.~Ballesteros and M.~Taoso, \emph{{Primordial black hole dark matter from
  single field inflation}},
  \href{https://doi.org/10.1103/PhysRevD.97.023501}{\emph{Phys. Rev. D}
  {\bfseries 97} (2018) 023501}
  [\href{https://arxiv.org/abs/1709.05565}{{\ttfamily 1709.05565}}].

\bibitem{Carr:1975qj}
B.J.~Carr, \emph{{The Primordial black hole mass spectrum}},
  \href{https://doi.org/10.1086/153853}{\emph{Astrophys. J.} {\bfseries 201}
  (1975) 1}.

\bibitem{Laha:2019ssq}
R.~Laha, \emph{{Primordial Black Holes as a Dark Matter Candidate Are Severely
  Constrained by the Galactic Center 511 keV $\gamma$ -Ray Line}},
  \href{https://doi.org/10.1103/PhysRevLett.123.251101}{\emph{Phys. Rev. Lett.}
  {\bfseries 123} (2019) 251101}
  [\href{https://arxiv.org/abs/1906.09994}{{\ttfamily 1906.09994}}].

\bibitem{Dasgupta:2019cae}
B.~Dasgupta, R.~Laha and A.~Ray, \emph{{Neutrino and positron constraints on
  spinning primordial black hole dark matter}},
  \href{https://doi.org/10.1103/PhysRevLett.125.101101}{\emph{Phys. Rev. Lett.}
  {\bfseries 125} (2020) 101101}
  [\href{https://arxiv.org/abs/1912.01014}{{\ttfamily 1912.01014}}].

\bibitem{Carr:2020gox}
B.~Carr, K.~Kohri, Y.~Sendouda and J.~Yokoyama, \emph{{Constraints on
  primordial black holes}},
  \href{https://doi.org/10.1088/1361-6633/ac1e31}{\emph{Rept. Prog. Phys.}
  {\bfseries 84} (2021) 116902}
  [\href{https://arxiv.org/abs/2002.12778}{{\ttfamily 2002.12778}}].

\bibitem{Laha:2020ivk}
R.~Laha, J.B.~Mu\~noz and T.R.~Slatyer, \emph{{INTEGRAL constraints on
  primordial black holes and particle dark matter}},
  \href{https://doi.org/10.1103/PhysRevD.101.123514}{\emph{Phys. Rev. D}
  {\bfseries 101} (2020) 123514}
  [\href{https://arxiv.org/abs/2004.00627}{{\ttfamily 2004.00627}}].

\bibitem{Ray:2021mxu}
A.~Ray, R.~Laha, J.B.~Mu\~noz and R.~Caputo, \emph{{Near future MeV telescopes
  can discover asteroid-mass primordial black hole dark matter}},
  \href{https://doi.org/10.1103/PhysRevD.104.023516}{\emph{Phys. Rev. D}
  {\bfseries 104} (2021) 023516}
  [\href{https://arxiv.org/abs/2102.06714}{{\ttfamily 2102.06714}}].

\bibitem{Saha:2021pqf}
A.K.~Saha and R.~Laha, \emph{{Sensitivities on nonspinning and spinning
  primordial black hole dark matter with global 21-cm troughs}},
  \href{https://doi.org/10.1103/PhysRevD.105.103026}{\emph{Phys. Rev. D}
  {\bfseries 105} (2022) 103026}
  [\href{https://arxiv.org/abs/2112.10794}{{\ttfamily 2112.10794}}].

\bibitem{Alexandre:2024nuo}
A.~Alexandre, G.~Dvali and E.~Koutsangelas, \emph{{New mass window for
  primordial black holes as dark matter from the memory burden effect}},
  \href{https://doi.org/10.1103/PhysRevD.110.036004}{\emph{Phys. Rev. D}
  {\bfseries 110} (2024) 036004}
  [\href{https://arxiv.org/abs/2402.14069}{{\ttfamily 2402.14069}}].

\bibitem{Thoss:2024hsr}
V.~Thoss, A.~Burkert and K.~Kohri, \emph{{Breakdown of Hawking Evaporation
  opens new Mass Window for Primordial Black Holes as Dark Matter Candidate}},
  \href{https://arxiv.org/abs/2402.17823}{{\ttfamily 2402.17823}}.

\bibitem{Dvali:2020wft}
G.~Dvali, L.~Eisemann, M.~Michel and S.~Zell, \emph{{Black hole metamorphosis
  and stabilization by memory burden}},
  \href{https://doi.org/10.1103/PhysRevD.102.103523}{\emph{Phys. Rev. D}
  {\bfseries 102} (2020) 103523}
  [\href{https://arxiv.org/abs/2006.00011}{{\ttfamily 2006.00011}}].

\bibitem{Dvali:2018ytn}
G.~Dvali, L.~Eisemann, M.~Michel and S.~Zell, \emph{{Universe's Primordial
  Quantum Memories}},
  \href{https://doi.org/10.1088/1475-7516/2019/03/010}{\emph{JCAP} {\bfseries
  03} (2019) 010} [\href{https://arxiv.org/abs/1812.08749}{{\ttfamily
  1812.08749}}].

\bibitem{Hamaide:2023ayu}
L.~Hamaide, L.~Heurtier, S.-Q.~Hu and A.~Cheek, \emph{{Primordial black holes
  are true vacuum nurseries}},
  \href{https://doi.org/10.1016/j.physletb.2024.138895}{\emph{Phys. Lett. B}
  {\bfseries 856} (2024) 138895}
  [\href{https://arxiv.org/abs/2311.01869}{{\ttfamily 2311.01869}}].

\bibitem{Virbhadra:1999nm}
K.S.~Virbhadra and G.F.R.~Ellis, \emph{{Schwarzschild black hole lensing}},
  \href{https://doi.org/10.1103/PhysRevD.62.084003}{\emph{Phys. Rev. D}
  {\bfseries 62} (2000) 084003}
  [\href{https://arxiv.org/abs/astro-ph/9904193}{{\ttfamily
  astro-ph/9904193}}].

\bibitem{Mroz:2024mse}
P.~Mr\'oz et~al., \emph{{No massive black holes in the Milky Way halo}},
  \href{https://doi.org/10.1038/s41586-024-07704-6}{\emph{Nature} {\bfseries
  632} (2024) 749} [\href{https://arxiv.org/abs/2403.02386}{{\ttfamily
  2403.02386}}].

\end{thebibliography}\endgroup

\end{document}